\pdfoutput=1
\documentclass{article}

\usepackage{iclr2027_conference,times}
\iclrfinalcopy

\usepackage{microtype}
\usepackage{amsmath,amssymb}
\usepackage{booktabs}
\usepackage{placeins}
\usepackage{needspace}
\usepackage{array}
\usepackage{multirow}
\usepackage{graphicx}
\usepackage{algorithm}
\usepackage{algpseudocode}
\usepackage{hyperref}
\usepackage{url}
\usepackage{fontawesome5}
\usepackage[capitalize,noabbrev]{cleveref}
\definecolor{darkblue}{rgb}{0.0,0.0,0.66}
\hypersetup{colorlinks=true,linkcolor=darkblue,citecolor=darkblue}

\newcommand{\laribo}{La-Ribo}
\newcommand{\laribobase}{La-Ribo-Base}
\newcommand{\laribotri}{La-Ribo-Tri}
\newcommand{\laribotrithree}{La-Ribo-Tri3}
\newcommand{\laribotrifour}{La-Ribo-Tri4}

\newcommand{\rnff}{RNA-FrameFlow}
\newcommand{\riboflowmix}{RiboFlow-Mix}
\newcommand{\riboflowexp}{RiboFlow-Exp}
\newcommand{\grnade}{gRNAde}
\newcommand{\rhofold}{RhoFold}

\newcommand{\bothfour}{\mathrm{both@4.0}}
\newcommand{\bothfive}{\mathrm{both@5.5}}
\newcommand{\tmfortyfive}{\mathrm{TM@0.45}}
\newcommand{\meanstd}[2]{\ensuremath{#1{\scriptstyle\,\pm\,#2}}}
\newcommand{\bestmeanstd}[2]{\ensuremath{\mathbf{#1}{\scriptstyle\,\pm\,#2}}}

\crefname{appendix}{Appendix}{Appendices}
\Crefname{appendix}{Appendix}{Appendices}
\crefname{subappendix}{Appendix}{Appendices}
\Crefname{subappendix}{Appendix}{Appendices}
\crefname{algorithm}{Algorithm}{Algorithms}
\Crefname{algorithm}{Algorithm}{Algorithms}
\newcommand{\setupappendixrefs}{%
  \crefalias{section}{appendix}%
  \crefalias{subsection}{subappendix}%
  \crefalias{subsubsection}{subsubappendix}%
}
\newcommand{\appentry}[1]{%
  \noindent\hyperref[#1]{\textbf{\ref*{#1}\quad\nameref*{#1}}}%
  \dotfill\hyperref[#1]{\pageref*{#1}}\par\smallskip
}
\newcommand{\appsub}[1]{%
  \noindent\hspace{1em}\hyperref[#1]{\ref*{#1}\quad\nameref*{#1}}%
  \dotfill\hyperref[#1]{\pageref*{#1}}\par
}

\title{La-Ribo: RNA Co-Design\\via Geometry--Latent Flow Matching}

\author{Runze Ma$^{1}$ \quad Will Hua$^{2}$ \quad Shuangjia Zheng$^{1}$\thanks{Corresponding author.} \\[3pt]
$^{1}$Shanghai Jiao Tong University \quad $^{2}$Independent Researcher \\[3pt]
\texttt{\{runze.ma, shuangjia.zheng\}@sjtu.edu.cn}}

\begin{document}

\maketitle
\lhead{Preprint}

\begin{center}\vspace{-0.27in}\small
\begin{tabular}{@{}c@{\hspace{0.5em}}l@{}}
\faGithub & : \url{https://github.com/GENTEL-lab/La-Ribo}\\
\includegraphics[height=1.05em]{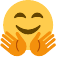} & : \url{https://huggingface.co/GENTEL-Lab/La-Ribo}
\end{tabular}\vspace{0.06in}
\end{center}

\begin{abstract}

RNA function arises from the coupling of nucleotide sequence and three-dimensional structure, motivating their joint design. Coordinating global folding with nucleotide-level detail remains challenging under limited structural supervision. We introduce La-Ribo, a generative framework for RNA sequence--structure co-design via geometry--latent flow matching. La-Ribo retains a sparse phosphate--sugar--base scaffold and encodes nucleotide identity and local conformation in residue-wise latents. A shared flow network generates both jointly, and an RNA-specific decoder then reconstructs all heavy atoms. To expand supervision, we construct a quality-controlled corpus of 168,561 RNA structures, integrating experimental data with predictions from three folding models, including 10,631 MSA-supported structures generated in this work. La-Ribo improves designability and codesignability over the evaluated baselines across sampling budgets and two refolding models, and the same prior supports scaffold-conditioned inverse folding without additional training.

\end{abstract}

\section{Introduction}
\label{sec:intro}

Ribonucleic acid (RNA) performs diverse functions, including catalysis, gene regulation, and molecular recognition, and these functions depend jointly on nucleotide sequence and three-dimensional structure~\citep{wilson2021potential,cao2024identification,childs2022targeting}. Computational design has traditionally approached these modalities separately. Sequence generators propose nucleotide sequences for downstream structural assessment, while backbone generators typically obtain sequences through inverse folding and assess compatibility by refolding~\citep{anand2024rnaframeflow}. Recent co-design methods instead generate sequence and structure together, enabling direct exploration of RNA sequence--structure space~\citep{rubin2025ribogen,ma2026riboflow,li2026joint}. Coordinating nucleotide identity with global folding nevertheless remains difficult, particularly under the limited structural supervision available for RNA.

RNA co-design faces two challenges: representation and structural supervision. A useful representation must preserve long-range tertiary organization while capturing nucleotide-specific local conformations. Partially latent protein models offer a complementary strategy, retaining coarse geometry explicitly while absorbing sequence and local detail into continuous per-residue latents~\citep{geffner2026laproteina}. Transferring this strategy to RNA is not direct. In proteins, the backbone anchors that define a residue frame are themselves a rigid unit; in RNA, anchors that span the phosphate, sugar, and base groups are not, so informative geometric anchors cannot serve directly as the frame for atom construction. Structural supervision is the second constraint. Experimental RNA structures number only in the thousands, far fewer than those available for proteins, so recent work supplements them with predicted structures. Predictions from different folding models differ in confidence, coverage, and length distribution, however, and they can transfer the biases of the folding models used to produce them, which makes consistent quality control necessary.

We introduce La-Ribo, a generative framework for RNA sequence--structure co-design via geometry--latent flow matching. La-Ribo retains $\mathrm{P}$, $\mathrm{C4}'$, and glycosidic $\mathrm{N}$ as sparse geometric anchors, while a scaffold-conditioned autoencoder learns residue-wise latents encoding nucleotide identity and local conformation. Nucleotide identity is absorbed into these continuous latents, so generation proceeds entirely in a continuous space and requires no discrete--continuous hybrid flow. The decoder predicts a residual rotation from the scaffold-defined frame to a canonical sugar frame and reconstructs heavy atoms using nucleotide templates and torsions; this learned interface separates the choice of informative anchors from the rigid-frame requirements of atom construction. A shared geometry--latent flow subsequently learns the joint distribution of scaffold coordinates and latents, from which the frozen decoder recovers sequences and structures. With geometry kept explicit, fixing the observed scaffold and integrating only the latent modality turns the same trained model into a scaffold-conditioned inverse-folding model, without conditional training.

To expand structural supervision, we construct a quality-controlled corpus of 168,561 RNA structures, combining experimental entries with predictions from OpenFold3, NuFold, and trRosettaRNA2~\citep{adamczyk2022rnasolo,shen2024accurate,openfold3preview,kagaya2025nufold,wang2026trrosettarna2}. Beyond curating existing resources, we generate 10,631 trRosettaRNA2 structures with multiple sequence alignments, confidence filtering, and structural relaxation. As predicted structures vary in reliability across sources, we apply source-specific confidence screening and common eligibility criteria before unified preprocessing.

Our main contributions are as follows. (\romannumeral 1) \textbf{Representation:} We introduce an RNA-specific geometry--latent representation combining sparse $\mathrm{P}$, $\mathrm{C4}'$, and glycosidic $\mathrm{N}$ anchors with residue-wise latents, linked by a learned scaffold-to-sugar frame transformation. (\romannumeral 2) \textbf{Generation:} A shared flow network generates geometry and latents jointly, and with geometry kept explicit, the same trained prior serves as a scaffold-conditioned inverse-folding model without conditional training. (\romannumeral 3) \textbf{Dataset:} We construct a quality-controlled corpus of 168,561 RNA structures integrating experimental and predicted sources under source-specific screening, including 10,631 MSA-supported and relaxed trRosettaRNA2 structures generated in this work. (\romannumeral 4) \textbf{Results:} La-Ribo improves designability and codesignability over the evaluated baselines, across sampling budgets and refolding models.

\section{Related work}
\label{sec:related}

\subsection{RNA generation, inverse folding, and co-design}
Computational RNA design spans sequence generation, inverse folding, structure generation, and joint sequence--structure generation. Sequence-centric models learn distributions over RNA sequences with language modeling or diffusion, extending \textit{de novo} generation toward function-, family-, and structure-conditioned design~\citep{zhang2025rnagenesis,patel2025evoflowrna,wang2026rdiffusion}. Structure-centric methods instead generate RNA geometry: \rnff{} applies flow matching to unconditional backbone generation, with sequences recovered through inverse folding~\citep{anand2024rnaframeflow}, while RNAbpFlow generates sequence- and base-pair-conditioned all-atom ensembles for conformational sampling rather than \textit{de novo} design~\citep{tarafder2026rnabpflow}. Inverse-folding models generate sequences for a target structure~\citep{joshi2024grnade,kubaney2025nampnn,huang2024ribodiffusion}.

Co-design methods generate both modalities within one generative process. RNAFlow integrates inverse folding with a pre-trained structure model for protein-conditioned design~\citep{nori2024rnaflow}, and RiboFlow pretrains an unconditional co-design model before fine-tuning it for target-molecule conditioning~\citep{ma2026riboflow}. Among unconditional methods, RiboGen combines discrete and continuous flows over nucleotide identities and all-atom geometry~\citep{rubin2025ribogen}, Joint Diffusion couples sequence and structure through a pre-trained RoseTTAFold2NA denoiser~\citep{li2026joint}, and DuetRNA introduces a base--sugar dual-frame representation~\citep{li2026duetrna}. \laribo{} instead pairs a sparse geometric state with residue-wise continuous latents. As nucleotide identity is carried by the latents rather than modeled as a discrete variable, generation proceeds as a single continuous flow rather than a discrete--continuous hybrid, and the same prior is reusable for scaffold-conditioned inverse folding without separate conditional training.

\subsection{Explicit and latent biomolecular representations}
Representations for biomolecular generation must capture global structure and local chemical detail. Geometric models commonly represent structures using Cartesian coordinates, rigid frames, torsions, or combinations thereof, while multimodal approaches couple these geometric states with sequence information~\citep{yim2023fast,bose2024se3,campbell2024generative,huguet2024sequence}. Latent generative models provide a complementary strategy by compressing high-dimensional observations before applying diffusion or flow matching~\citep{kingma2013auto,rombach2022high,esser2024scaling}. Recent biomolecular models increasingly incorporate learned representations into geometric modeling. La-Proteina introduces a partially latent representation for atomistic protein generation, retaining coarse coordinates explicitly while absorbing sequence and local atomic detail into continuous per-residue latents~\citep{geffner2026laproteina}. This strategy does not transfer directly to RNA, where informative geometric anchors span the phosphate, sugar, and base groups and therefore do not form the rigid unit that atom construction requires. For RNA, RiboSphere learns discrete structural tokens through geometric quantization and transfers the learned representation to reconstruction and downstream tasks such as inverse folding~\citep{zhang2026ribosphere}. \laribo{} instead pairs sparse RNA geometry with continuous residue-wise latents, connected by a learned scaffold-to-sugar frame transformation that resolves this mismatch.

\section{Method}
\label{sec:method}

\begin{figure}[t]
    \centering
    \includegraphics[width=0.95\linewidth]{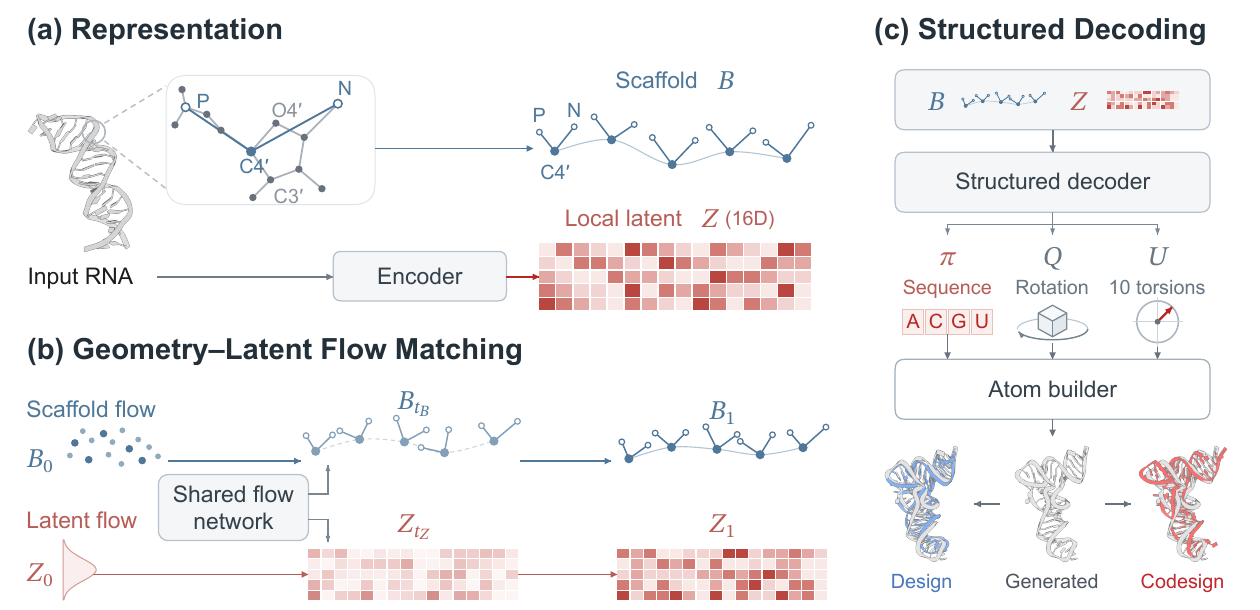}
    \caption{Overview of \laribo{}.
    (a) Each residue is represented by an explicit P/C4'/glycosidic-N
    scaffold (blue) and an encoded 16-dimensional latent (red).
    (b) A shared flow network jointly generates states with
    modality-specific schedules.
    (c) The frozen decoder predicts nucleotide identity, a scaffold-to-sugar
    rotation, and torsions, from which the atom builder builds all heavy atoms.}
    \label{fig:overview}
\end{figure}

\laribo{} first learns an RNA-specific geometry--latent representation through a geometry-aware autoencoder, and then models the resulting scaffold and latent states through geometry--latent flow matching. \Cref{fig:overview} summarizes the representation, generative process, and heavy-atom reconstruction.

\subsection{Geometry--latent representation}
\label{sec:representation}

We consider a canonical RNA chain of length \(N\), with sequence
\(S=(s_1,\ldots,s_N)\), where \(s_i\in\mathcal{A}=\{A,C,G,U\}\).
Following the commonly used atom37 convention in biomolecular
modeling~\citep{jumper2021highly,geffner2026laproteina}, we represent its
structure by heavy-atom coordinates
\(X\in\mathbb{R}^{N\times37\times3}\).
Let \(x_i^a\in\mathbb{R}^3\) denote the coordinate of atom \(a\) in residue \(i\). Rather than directly generating the complete atomistic state, we retain a
sparse geometric scaffold
\begin{equation}
b_i=
\left[
x_i^{\mathrm{P}};
x_i^{\mathrm{C4^\prime}};
x_i^{\nu_i}
\right]
\in\mathbb{R}^{d_B},
\qquad
d_B=9,
\label{eq:scaffold}
\end{equation}
where \(\nu_i=\mathrm{N9}\) for \(s_i\in\{A,G\}\) and
\(\nu_i=\mathrm{N1}\) for \(s_i\in\{C,U\}\).
Thus, each scaffold state contains the coordinates of
\(\mathrm{P}\), \(\mathrm{C4^\prime}\), and the glycosidic nitrogen.
The scaffold \(B=(b_1,\ldots,b_N)\) captures phosphate, ribose, and
base-attachment geometry. The
\(\mathrm{P}\)--\(\mathrm{C4^\prime}\) trace provides a reduced description
of RNA backbone conformation through standard pseudotorsional
geometry~\citep{wadley2007evaluating}, while the glycosidic nitrogen provides
an explicit anchor for the base attachment. \Cref{app:representation_ablation} compares one-, two-, and three-anchor scaffolds and latent widths, supporting this choice.

Nucleotide identity and the local atomic detail not determined by the scaffold, including non-anchor heavy-atom positions, ribose conformation, and base orientation, are represented by residue-wise latent variables \(z_i\in\mathbb{R}^{d_Z}\), with \(Z=(z_1,\ldots,z_N)\) and \(d_Z=16\) unless stated otherwise. \laribo{} therefore represents each residue by a \((d_B+d_Z)\)-dimensional geometry--latent state, comprising 9 explicit scaffold coordinates and 16 learned latent dimensions.

The latent representation is learned by an encoder \(E_\phi\), and the
sequence and atomistic structure are reconstructed by a structured decoder
\(D_\psi\), where \(\phi\) and \(\psi\) denote their learnable parameters.
\Cref{sec:stage1} describes this autoencoding process, while
\cref{sec:stage2} describes learning generative dynamics over the joint state
\((B,Z)\).

\subsection{Scaffold-conditioned autoencoding}
\label{sec:stage1}

\paragraph{Latent encoding.}
We learn the residue-wise latent representation using an encoder \(E_\phi\).
Given the sequence \(S\), structure \(X\), and scaffold \(B\), the
encoder predicts
\begin{equation}
(\mu,\sigma)=E_\phi(S,X,B),
\qquad
z_i=\mu_i+\sigma_i\odot\epsilon_i,
\quad
\epsilon_i\sim\mathcal N(0,I),
\label{eq:latent_encoding}
\end{equation}
where \(\mu,\sigma\in\mathbb{R}^{N\times d_Z}\) are the latent means and standard deviations. Although
\(Z=(z_1,\ldots,z_N)\) is residue-wise, the context-aware encoder allows each
latent to depend on the full RNA chain.

\paragraph{Frame-aware decoding.}
The geometric scaffold and atom builder have different representational
requirements. The \(\mathrm{P}/\mathrm{C4^\prime}/\mathrm{N}\) scaffold
captures backbone and base-attachment geometry, but its three anchors belong
to different chemical groups rather than a single rigid unit. Atomistic reconstruction instead uses a canonical frame within the
ribose, constructed from
\(\mathrm{C4^\prime}/\mathrm{O4^\prime}/\mathrm{C3^\prime}\) following
\rnff{}~\citep{anand2024rnaframeflow}. These two requirements select different atoms: \(\mathrm{O4^\prime}\) and \(\mathrm{C3^\prime}\) are confined to the ribose ring and carry little information about backbone trace or base placement, whereas \(\mathrm{P}\) and the glycosidic nitrogen do. We therefore keep the informative anchors in the generated state and learn the transformation between the two frames. Writing \(\mathcal F(o,u,v)\in\mathrm{SO}(3)\) for the orthonormal local frame centered at \(o\), we define
\begin{equation}
R_i^p
=
\mathcal F
(x_i^{\mathrm{C4^\prime}},x_i^{\nu_i},x_i^{\mathrm{P}}),
\qquad
R_i^c
=
\mathcal F
(x_i^{\mathrm{C4^\prime}},x_i^{\mathrm{O4^\prime}},x_i^{\mathrm{C3^\prime}}),
\label{eq:scaffold_sugar_frames}
\end{equation}
where \(R_i^p\) is the scaffold pseudo-frame and \(R_i^c\) is the canonical
sugar frame. Given \((B,Z)\), the structured decoder predicts
\begin{equation}
(\pi_i,Q_i,U_i)=D_\psi(B,Z)_i,
\qquad
R_i^{\mathrm{dec}}=R_i^p Q_i .
\label{eq:decoder_outputs}
\end{equation}
Here \(\pi_i\) is a probability distribution over
\(\mathcal A=\{A,C,G,U\}\), \(Q_i\in\mathrm{SO}(3)\) is the predicted
rotation from the scaffold pseudo-frame to the sugar frame, and
\(U_i=\{u_{ik}\}_{k=1}^{10}\) contains 10 normalized sine--cosine torsion
representations. The decoded frame \(R_i^{\mathrm{dec}}\), torsions \(U_i\),
and nucleotide identity specified by \(\pi_i\) together parameterize the
structured atom builder for heavy-atom reconstruction. The construction of \(\mathcal F\) and the torsion and atom-construction conventions are provided in \cref{app:exact_losses}.

\paragraph{Training objective.}
The autoencoder jointly supervises nucleotide identity, the
scaffold-to-sugar frame transformation, torsional geometry, and atomistic
reconstruction. We write the geometric reconstruction objective under
template view \(v\) as
\begin{equation}
\mathcal L_{\mathrm{geom}}^{v}
=
\mathcal L_{\mathrm{atom}}^{v}
+
\mathcal L_{\mathrm{scaf}}^{v}
+
\mathcal L_{\mathrm{pseudo}}^{v},
\qquad
v\in\{\mathrm{gt},\mathrm{pred}\},
\label{eq:geometry_ae_loss}
\end{equation}
and optimize
\begin{equation}
\mathcal L_{\mathrm{AE}}
=
\mathcal L_{\mathrm{seq}}
+
\mathcal L_{\mathrm{rot}}
+
\mathcal L_{\mathrm{tors}}
+
\lambda_{\mathrm{gt}}\mathcal L_{\mathrm{geom}}^{\mathrm{gt}}
+
\lambda_{\mathrm{pred}}\mathcal L_{\mathrm{geom}}^{\mathrm{pred}}
+
\lambda_{\mathrm{KL}}\mathcal L_{\mathrm{KL}} .
\label{eq:vae_loss}
\end{equation}
The sequence, rotation, and torsion terms supervise nucleotide identity,
sugar-frame orientation, and local conformation; the geometric terms compare
heavy atoms, scaffold coordinates, and backbone pseudotorsions.
The ground-truth template provides identity-consistent supervision, while
the predicted-template term evaluates reconstruction with the inference-time
template-selection rule. The KL term regularizes the latent distribution.
Exact losses, weights, and masks are in
\cref{app:exact_losses}.

\subsection{Geometry--latent flow matching}
\label{sec:stage2}

With the encoder \(E_\phi\) and decoder \(D_\psi\) fixed, we learn generative dynamics over \((B,Z)\) through flow matching~\citep{lipmanflow,liuflow}. Scaffold coordinates are centered at the valid \(\mathrm{C4^\prime}\) centroid and globally scaled, while posterior-mean latents are standardized dimension-wise, yielding
\begin{equation}
x_1^B\in\mathbb{R}^{N\times d_B},
\qquad
x_1^Z\in\mathbb{R}^{N\times d_Z}.
\label{eq:clean_states}
\end{equation}

Let \(a\in\{B,Z\}\) index the scaffold and latent modalities. For each
modality, we pair the clean state \(x_1^a\) with an independently sampled Gaussian base state \(x_0^a\), centered in the same way for the scaffold modality, and define the linear path
\begin{equation}
x_{t_a}^a=(1-t_a)x_0^a+t_a x_1^a,
\qquad
a\in\{B,Z\}.
\label{eq:geometry_latent_path}
\end{equation}
The scaffold and latent times \(t_B\) and \(t_Z\) are sampled independently,
allowing the two modalities to occupy different noise levels during training.

A shared velocity network, parameterized by \(\theta\), predicts both
components \(v_\theta=(v_\theta^B,v_\theta^Z)\) from
\(x_t=(x_{t_B}^B,x_{t_Z}^Z)\). Here \(v_\theta^B\) includes scaffold
recentering. Throughout the main text we write \(v_\theta(x_t,t_B,t_Z)\), omitting the self-conditioning input \(x_{\mathrm{sc}}\) from the argument list; \cref{app:flow_details} gives the full signature. We optimize
\begin{equation}
\mathcal L_{\mathrm{FM}}
=
\sum_{a\in\{B,Z\}}
\mathbb E\!\left[
\frac{1}{d_a|\mathcal R|}
\sum_{i\in\mathcal R}
\left\|
v_\theta^a(x_t,t_B,t_Z)_i
-
(x_1^a-x_0^a)_i
\right\|_2^2
\right],
\label{eq:flow_loss}
\end{equation}
where \(\mathcal R\) contains valid residues. Training uses only geometry--latent targets, without decoded-coordinate or sequence losses. The decoder is frozen at this stage, so the flow is fit directly in the representation space rather than through a reconstruction pathway, which keeps the two stages independent and avoids backpropagating through the non-differentiable template selection of \cref{sec:stage1}.
Exact normalization, scaffold recentering, time sampling, self-conditioning,
and endpoint stabilization are described in \cref{app:flow_details}.

\subsection{Sampling}
\label{sec:sampling}

For unconditional generation, scaffold and latent states are initialized from
their base distributions and integrated jointly with forward Euler using
modality-specific schedules,
\begin{equation}
t_k^a
=
\left(\frac{k}{K}\right)^{\rho_a},
\qquad
a\in\{B,Z\},
\qquad
k=0,\ldots,K,
\label{eq:sampling_schedule}
\end{equation}
where \(K\) is the number of Euler steps and \(\rho_B,\rho_Z\) control the
scaffold and latent schedules, respectively. After reversing normalization, the terminal state \((B,Z)\) is decoded by \(D_\psi\) into a nucleotide sequence and full heavy-atom structure. For scaffold-conditioned inverse folding, the observed scaffold is fixed at \(t_B=1\) and only the latent modality is integrated. The resulting latent state is decoded with the supplied scaffold. Training times are restricted to \([0.01,0.99]\) (\cref{app:flow_details}), so \(t_B=1\) slightly extrapolates beyond the training range; the procedure is nevertheless approximate, as holding one modality fixed does not exactly recover the conditional of the learned joint distribution.

\section{Experiments}
\label{sec:experiments}

\subsection{Experimental setup}
\label{sec:data}

\suppressfloats[t]
\begin{table}[!t]
\centering
\caption{Unconditional RNA generation at 50--400 steps. Mean $\pm$ standard deviation over three 600-RNA campaigns; bold marks the best mean at each step count.}
\vspace{0.4\baselineskip}
\label{tab:main}
\footnotesize
\setlength{\tabcolsep}{3pt}
\begin{tabular}{@{}lr rrr rrr@{}}
\toprule
& & \multicolumn{3}{c}{Designability} & \multicolumn{3}{c}{Codesignability}\\
\cmidrule(lr){3-5}\cmidrule(l){6-8}
Method & Steps & $\bothfour\uparrow$ & $\tmfortyfive\uparrow$ & \#Clust.$\uparrow$ & $\bothfour\uparrow$ & $\tmfortyfive\uparrow$ & \#Clust.$\uparrow$\\
\midrule
\multirow{4}{*}{\rnff{}} & 50 & \meanstd{25.17}{1.09} & \meanstd{36.61}{0.19} & \meanstd{24.00}{1.73} & -- & -- & -- \\
 & 100 & \meanstd{26.61}{1.18} & \meanstd{39.67}{0.33} & \meanstd{27.00}{2.65} & -- & -- & -- \\
 & 200 & \meanstd{26.17}{1.76} & \meanstd{38.44}{1.84} & \meanstd{26.00}{1.00} & -- & -- & -- \\
 & 400 & \meanstd{26.89}{1.69} & \meanstd{38.89}{1.69} & \meanstd{27.00}{3.61} & -- & -- & -- \\
\midrule
\multirow{4}{*}{\riboflowexp{}} & 50 & \meanstd{22.28}{0.63} & \meanstd{30.44}{0.54} & \meanstd{19.33}{4.04} & \meanstd{19.22}{0.35} & \meanstd{24.94}{0.10} & \meanstd{15.33}{2.52} \\
 & 100 & \meanstd{22.22}{0.54} & \meanstd{31.22}{0.98} & \meanstd{18.67}{1.53} & \meanstd{19.28}{0.42} & \meanstd{25.78}{1.25} & \meanstd{15.67}{2.08} \\
 & 200 & \meanstd{20.50}{0.73} & \meanstd{30.06}{0.69} & \meanstd{18.00}{2.65} & \meanstd{19.33}{1.61} & \meanstd{26.44}{2.55} & \meanstd{15.67}{1.15} \\
 & 400 & \meanstd{21.39}{0.54} & \meanstd{29.94}{0.77} & \meanstd{17.00}{2.00} & \meanstd{20.06}{0.19} & \meanstd{27.61}{0.69} & \meanstd{15.67}{1.53} \\
\midrule
\multirow{4}{*}{\riboflowmix{}} & 50 & \meanstd{29.00}{0.17} & \meanstd{37.28}{1.09} & \meanstd{23.00}{3.46} & \meanstd{26.61}{1.34} & \meanstd{31.83}{0.58} & \meanstd{19.00}{2.65} \\
 & 100 & \meanstd{29.06}{0.54} & \meanstd{38.44}{0.70} & \meanstd{22.33}{2.31} & \meanstd{26.67}{0.67} & \meanstd{32.50}{0.71} & \meanstd{19.33}{2.08} \\
 & 200 & \meanstd{27.61}{1.07} & \meanstd{36.89}{0.92} & \meanstd{21.67}{1.53} & \meanstd{26.00}{1.38} & \meanstd{33.06}{1.13} & \meanstd{19.00}{1.73} \\
 & 400 & \meanstd{28.39}{0.10} & \meanstd{37.39}{0.91} & \meanstd{20.67}{2.31} & \meanstd{26.61}{1.57} & \meanstd{34.72}{1.34} & \meanstd{19.33}{1.53} \\
\midrule
\multirow{4}{*}{\laribobase{}} & 50 & \meanstd{36.11}{2.07} & \meanstd{50.78}{2.00} & \meanstd{22.67}{1.53} & \bestmeanstd{30.00}{2.77} & \bestmeanstd{52.78}{2.18} & \meanstd{24.00}{1.00} \\
 & 100 & \meanstd{37.11}{0.98} & \meanstd{51.83}{1.20} & \meanstd{22.67}{1.15} & \bestmeanstd{32.11}{2.14} & \meanstd{55.28}{1.55} & \meanstd{23.33}{3.06} \\
 & 200 & \meanstd{38.22}{1.27} & \meanstd{53.00}{0.67} & \meanstd{23.00}{1.73} & \bestmeanstd{32.56}{2.25} & \meanstd{56.72}{1.78} & \meanstd{25.33}{2.31} \\
 & 400 & \meanstd{37.72}{1.13} & \meanstd{53.11}{1.55} & \meanstd{24.33}{1.15} & \bestmeanstd{32.56}{1.58} & \meanstd{57.17}{0.29} & \meanstd{25.67}{3.51} \\
\midrule
\multirow{4}{*}{\laribotri{}} & 50 & \bestmeanstd{37.11}{0.25} & \bestmeanstd{54.06}{0.35} & \bestmeanstd{29.00}{2.65} & \meanstd{28.61}{1.13} & \meanstd{50.83}{2.09} & \bestmeanstd{28.00}{1.00} \\
 & 100 & \bestmeanstd{37.78}{1.92} & \bestmeanstd{56.89}{2.22} & \bestmeanstd{28.33}{5.51} & \meanstd{30.78}{1.26} & \bestmeanstd{56.11}{1.95} & \bestmeanstd{25.33}{3.21} \\
 & 200 & \bestmeanstd{39.11}{2.12} & \bestmeanstd{58.83}{1.48} & \bestmeanstd{27.67}{1.15} & \meanstd{32.06}{0.75} & \bestmeanstd{57.89}{1.71} & \bestmeanstd{27.33}{0.58} \\
 & 400 & \bestmeanstd{38.61}{2.46} & \bestmeanstd{59.33}{0.93} & \bestmeanstd{29.33}{3.79} & \meanstd{32.33}{0.58} & \bestmeanstd{58.11}{0.84} & \bestmeanstd{28.00}{1.00} \\
\bottomrule\end{tabular}\end{table}

\begin{figure}[t]
    \centering
    \includegraphics[width=0.92\linewidth]{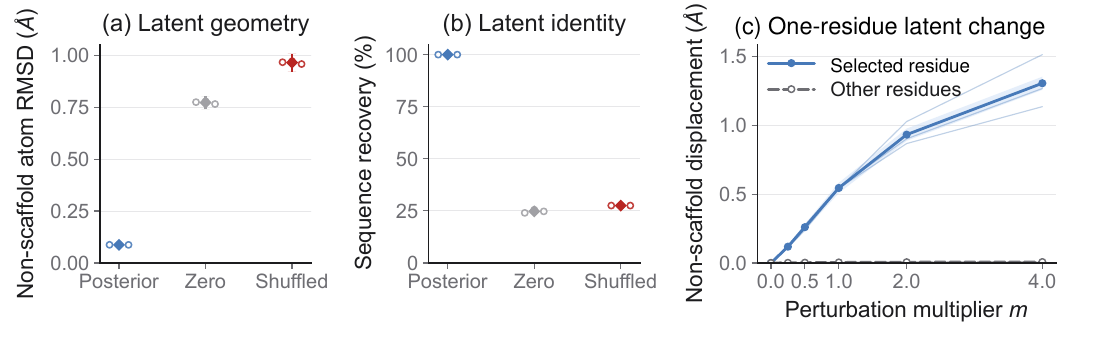}
    \caption{Latent interventions with a fixed scaffold. (a,b) compare posterior-mean, zero, and shuffled latents; (c) shows atom displacement from a separate one-residue latent perturbation.}
    \label{fig:vae_latent_interventions}
\end{figure}

\textbf{Dataset.}
We construct a corpus of 168,561 RNA structures from RNASolo, experimental
PDB chains included in the RhoFold release, predicted structures from
OpenFold3 and NuFold, and a trRosettaRNA2 set generated in this
work~\citep{adamczyk2022rnasolo,shen2024accurate,openfold3preview,
kagaya2025nufold,wang2026trrosettarna2}.
Entries span 16--256 nt with at least 90\% valid-frame coverage.
Sequences are globally clustered with MMseqs2 at 90\% identity and 95\%
coverage, and whole clusters are assigned approximately 95/5 to training
and validation, yielding 159,955 and 8,606 structures, respectively.
All learned-representation analyses and the inverse-folding
benchmark (\cref{sec:conditional_sequence}) use only the cluster-held-out
validation set, whose sequence clusters are disjoint from all training data.
\Cref{app:data_inventory} provides the complete data-construction workflow.

\textbf{Baselines and model variants.}
We compare against \rnff{}, evaluated with its released weights~\citep{anand2024rnaframeflow}, and RiboFlow~\citep{ma2026riboflow}, whose unconditional co-design stage we retrain in two regimes: \riboflowmix{} on the full \laribo{} training partition, and \riboflowexp{} on its experimental-only subset, matching the data regime of \rnff{}.
RiboGen~\citep{rubin2025ribogen}, Joint Diffusion~\citep{li2026joint}, and DuetRNA~\citep{li2026duetrna} had not released code at the time of evaluation, preventing matched reruns under our protocol.
Our two main variants share the same pretrained scaffold-conditioned autoencoder:
\laribobase{} keeps the pair representation fixed throughout the flow network,
whereas \laribotri{} refines it every two blocks using token-to-pair projections
and triangle multiplicative updates.
\Cref{app:config} provides the training details.

\subsection{Representation and latent analysis}
\label{sec:vae_eval}

\textbf{Setup.} To test what $Z$ contributes beyond the scaffold, we study two scaffold-conditioned autoencoders: VAE, trained on the training partition, and E-VAE, the same architecture trained only on experimental structures from the training partition. Both are evaluated on the same cluster-held-out validation set. With scaffold $B$ fixed, we decode three latent conditions: posterior ($z_i=\mu_i$), zero ($z_i=0$), and shuffle (a within-chain permutation of the $\mu_i$). A separate locality test changes one residue's latent while keeping $B$ and all other latents fixed. \textbf{Evaluation metrics.} Sequence recovery uses the decoder's nucleotide probabilities $\pi_i$. We report canonical sugar-frame angular error and non-scaffold heavy-atom RMSD; the latter excludes the scaffold anchors $x_i^{\mathrm P}$, $x_i^{\mathrm{C4^\prime}}$, and $x_i^{\nu_i}$. \Cref{tab:vae_latent_interventions} also reports reconstructed P and glycosidic-N errors. Locality compares atom displacement at the edited residue with all others as the perturbation magnitude $m$ grows. \textbf{Results.} Posterior latents reconstruct almost exactly: 100\% sequence recovery, sugar-frame error of $1.78^\circ$/$2.67^\circ$, and non-scaffold RMSD of 0.087/0.122~\AA{} for VAE and E-VAE, whose closeness indicates the corpus adds little at this stage. Zeroing the latents drops recovery to 24.77\%, chance level for four nucleotides, so the scaffold alone carries no identity information; RMSD rises to 0.773~\AA{}. Shuffling is worse still (0.965~\AA{}), showing the decoder acts on the supplied latents rather than an average conformation. Single-residue perturbations displace the edited residue's atoms far more than all others (\cref{fig:vae_latent_interventions}), so the latents are residue-local. One- and two-anchor scaffolds reach 0.159 and 0.130~\AA{} against 0.122~\AA{} for three anchors, at the same latent width and training regime (\cref{app:representation_ablation}). \Cref{app:vae_latent_controls} gives the complete metrics.

\begin{table}[!t]
\centering
\caption{Autoencoder corpus $\times$ flow architecture at 200 steps.
VAE is trained on the full corpus, E-VAE on its experimental subset; all four
flows use experimental structures only. Mean $\pm$ standard deviation over three
600-RNA campaigns; \rnff{} and \riboflowexp{} are references.}
\label{tab:stage2_vae_corpus_ablation}
\vspace{0.4\baselineskip}
\footnotesize\setlength{\tabcolsep}{6pt}
\begin{tabular}{@{}llrrrrr@{}}
\toprule
Autoencoder & Flow & $\bothfour\uparrow$ & $\bothfive\uparrow$ & $\tmfortyfive\uparrow$ & scRMSD $\downarrow$ & scTM $\uparrow$\\
\midrule
\multicolumn{2}{l}{\rnff{}} & \meanstd{26.17}{1.76} & \meanstd{31.56}{2.10} & \meanstd{38.44}{1.84} & \meanstd{11.47}{0.29} & \meanstd{0.413}{0.008}\\
\multicolumn{2}{l}{\riboflowexp{}} & \meanstd{20.50}{0.73} & \meanstd{25.11}{0.42} & \meanstd{30.06}{0.69} & \meanstd{12.73}{0.22} & \meanstd{0.408}{0.003}\\
\midrule
VAE & Base & \meanstd{25.78}{0.92} & \meanstd{29.83}{2.18} & \meanstd{33.67}{1.88} & \meanstd{13.51}{0.38} & \meanstd{0.374}{0.004}\\
VAE & Tri & \bestmeanstd{30.72}{1.39} & \bestmeanstd{36.17}{0.76} & \meanstd{43.44}{0.54} & \meanstd{11.74}{0.17} & \meanstd{0.433}{0.004}\\
E-VAE & Base & \meanstd{25.50}{1.20} & \meanstd{29.67}{1.33} & \meanstd{34.89}{1.13} & \meanstd{12.66}{0.14} & \meanstd{0.400}{0.002}\\
E-VAE & Tri & \meanstd{30.17}{1.36} & \meanstd{35.22}{1.11} & \bestmeanstd{44.06}{1.49} & \bestmeanstd{10.82}{0.15} & \bestmeanstd{0.444}{0.006}\\
\bottomrule
\end{tabular}\end{table}

\begin{figure}[!t]
    \centering
    \includegraphics[width=0.9\linewidth]{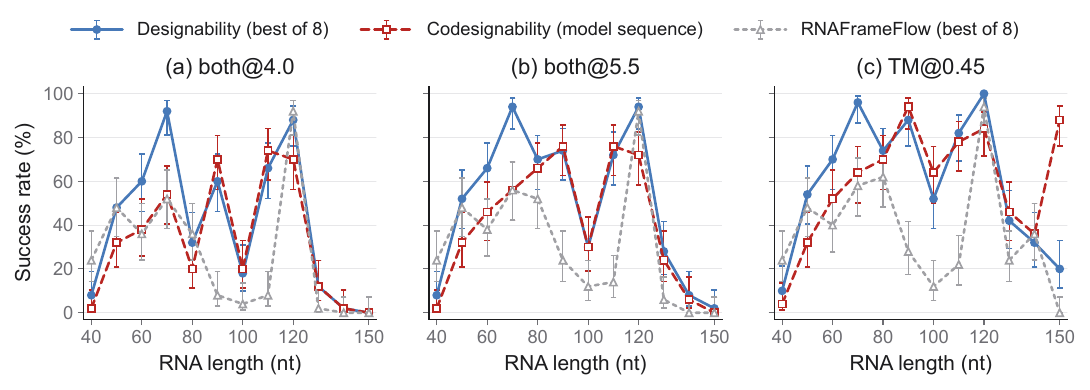}
    \caption{Length-resolved \laribotri{} success at 200 steps for one 600-RNA campaign (50 RNAs per length). Error bars show 95\% Wilson intervals.}
    \label{fig:length_resolved}
\end{figure}

\subsection{Unconditional RNA generation}
\label{sec:main_results}

\begin{figure}[!t]
    \centering
    \includegraphics[width=0.9\linewidth]{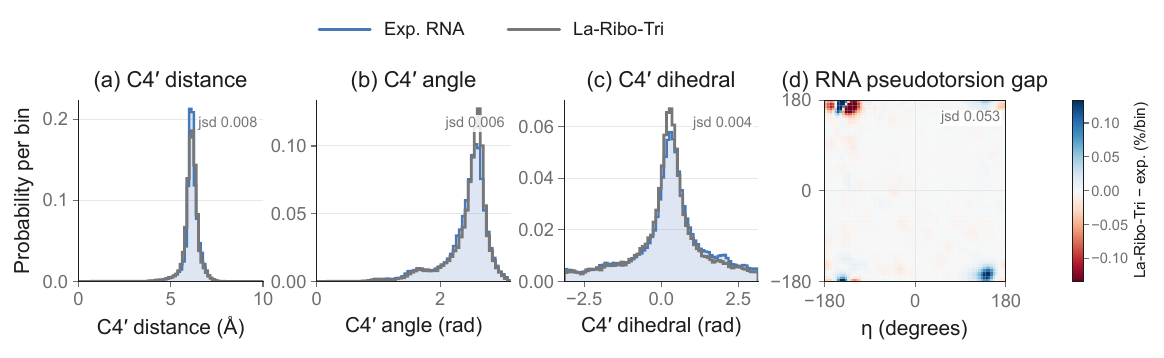}
    \caption{Local geometry of \laribotri{} (gray) and experimental RNA (blue; RNASolo and RhoFold PDB). (a--c) compare C4$'$ distance, angle, and dihedral; (d) shows generated-minus-experimental pseudotorsion density. JSD is in bits.}
    \label{fig:local_geometry_comparison}
\end{figure}

\textbf{Setup.} Following \rnff{} and RiboFlow~\citep{anand2024rnaframeflow,ma2026riboflow}, each campaign generates 600 RNAs, 50 at each length from 40 to 150 nt in 10-nt increments. We compare \rnff{}, \riboflowexp{} and \riboflowmix{}, \laribobase{}, and \laribotri{} at 50, 100, 200, and 400 steps, with three independently seeded campaigns per setting to estimate sampling variation. The La-Ribo variants use fixed modality schedules. \Cref{fig:local_geometry_comparison} compares local-geometry distributions from 200-step \laribotri{} campaigns with experimental RNA structures.

\textbf{Evaluation metrics.} Designability refolds eight \grnade{} sequences per generated structure and succeeds if any candidate passes; codesignability refolds the emitted sequence. Main-text scores use \rhofold{}; \cref{app:opendde_robustness} repeats the evaluation on the same structures and candidate sequences with OpenDDE. We align each refold with its generated structure using C4$'$ atoms to compute self-consistency RMSD (scRMSD) and TM-score (scTM). $\bothfour$ requires the same refold to have scRMSD$\leq4.0$~\AA{} and scTM$\geq0.45$; $\bothfive$ changes the RMSD cutoff to 5.5~\AA{}, and $\tmfortyfive$ requires scTM$\geq0.45$. \#Clust.\ counts qTMclust~\citep{zhang2022us} clusters among generated backbones whose refolds meet $\tmfortyfive$.

\begin{table}[!t]
\centering
\caption{Triangle-update interval at 200 steps. Interval counts blocks between pair updates; Base keeps pair features fixed. Mean $\pm$ standard deviation over three 600-RNA campaigns.}
\vspace{0.4\baselineskip}
\label{tab:triangle_frequency}
\footnotesize
\setlength{\tabcolsep}{2.5pt}
\begin{tabular}{@{}lc rrr rrr@{}}
\toprule
& & \multicolumn{3}{c}{Designability} & \multicolumn{3}{c}{Codesignability}\\
\cmidrule(lr){3-5}\cmidrule(l){6-8}
Model & Interval & $\bothfour\uparrow$ & $\bothfive\uparrow$ & $\tmfortyfive\uparrow$ & $\bothfour\uparrow$ & $\bothfive\uparrow$ & $\tmfortyfive\uparrow$\\
\midrule
\laribobase{} & --
& \meanstd{38.22}{1.27} & \meanstd{43.44}{0.48} & \meanstd{53.00}{0.67}
& \bestmeanstd{32.56}{2.25} & \bestmeanstd{41.28}{1.84} & \meanstd{56.72}{1.78} \\
\laribotrifour{} & 4
& \meanstd{37.33}{2.03} & \meanstd{45.00}{1.92} & \meanstd{55.33}{1.36}
& \meanstd{25.17}{0.59} & \meanstd{36.67}{1.21} & \meanstd{52.67}{1.59} \\
\laribotrithree{} & 3
& \meanstd{38.50}{2.09} & \meanstd{44.00}{1.89} & \meanstd{56.00}{1.41}
& \meanstd{26.83}{0.60} & \meanstd{36.50}{1.15} & \meanstd{52.00}{1.53} \\
\laribotri{} & 2
& \bestmeanstd{39.11}{2.12} & \bestmeanstd{48.39}{2.08} & \bestmeanstd{58.83}{1.48}
& \meanstd{32.06}{0.75} & \meanstd{39.94}{1.26} & \bestmeanstd{57.89}{1.71} \\
\bottomrule
\end{tabular}
\end{table}

\textbf{Results.} \laribotri{} leads strict designability ($\bothfour$) at all four step counts (\cref{tab:main}; \cref{tab:main_full} reports expanded metrics). At 200 steps it reaches $39.11\pm2.12\%$, over eleven points above the strongest baseline (\riboflowmix{} at $27.61\pm1.07\%$; \rnff{} reaches $26.17\pm1.76\%$); its strict codesignability is $32.06\pm0.75\%$ against $26.00\pm1.38\%$ for \riboflowmix{}. \riboflowexp{} falls below \rnff{} under matched experimental-only training, while \riboflowmix{} exceeds it. The OpenDDE evaluation favors both variants over their baselines (\cref{app:opendde_robustness}). The two variants are indistinguishable in codesignability, while \laribotri{} leads designability topology success and cluster coverage at every budget; \laribobase{} samples faster (\cref{app:sampling_factors}). Strict success degrades beyond 120\,nt (\cref{fig:length_resolved}), while local-geometry distributions track the experimental reference closely (JSD $\leq0.008$ bits for C4$'$ distance, angle, and dihedral; \cref{fig:local_geometry_comparison}). \textbf{Additional analyses.} Beyond self-consistency, the appendix evaluates secondary-structure agreement (\cref{app:secondary_agreement}), atomistic geometry including clashes and ribose pucker, where both La-Ribo variants improve on the baselines (\cref{app:atomistic}), and proximity to the training corpus (\cref{app:novelty}).

\subsection{Ablation studies}
\label{sec:ablations}

\paragraph{Autoencoder data and flow architecture}
\label{sec:ablation_autoencoder_flow}
We cross the corpus-trained VAE and experimental-only E-VAE with Base and Tri flows, all trained on the experimental partition and evaluated at 200 steps. Tri outperforms Base on all five metrics under both representations, with same-seed strict-designability gains of $4.94\pm0.54$ and $4.67\pm1.88$ points (\cref{tab:stage2_vae_corpus_ablation}). Representation and corpus each contribute. Trained only on experimental structures, E-VAE/Tri already reaches 30.17\%, above \rnff{} (26.17\%) and \riboflowexp{} (20.50\%), so the representation helps before any corpus expansion. With the autoencoder fixed, expanding the flow data adds 8.39 points for \laribo{} (39.11\% versus 30.72\%, \cref{tab:main}) and 7.11 for RiboFlow (27.61\%), so the gain is not representation-specific.

\paragraph{Triangle-update frequency}
\label{sec:ablation_triangle_frequency}
Holding the representation, corpus, and evaluation fixed, we compare triangle updates every two, three, or four blocks with fixed-pair Base. At 200 steps, strict designability is indistinguishable across the four variants (37.33--39.11\%), but topology success rises monotonically with update frequency (53.00\% for Base to 58.83\% for the two-block variant), and the two-block variant exceeds the sparser ones by more than five points in strict codesignability; Base remains comparable there at 32.56\% (\cref{tab:triangle_frequency}). \Cref{app:triangle_frequency} reports the 50--400-step results.

\begin{table}[!t]
\centering
\caption{Inverse folding on 1,002 experimental validation scaffolds, mean $\pm$ sample standard deviation over three independently seeded evaluations. Mean-of-eight averages candidates; best-of-eight uses one candidate for strict success. Rates are percentages; Unique counts distinct candidates (of eight); scRMSD is in \AA{}.}
\vspace{0.4\baselineskip}
\label{tab:conditional_sequence_main}
\scriptsize
\setlength{\tabcolsep}{1.2pt}
\begin{tabular*}{\linewidth}{@{\extracolsep{\fill}}lrrrrrrrr@{}}
\toprule
& \multicolumn{2}{c}{Sequence statistics} & \multicolumn{3}{c}{Mean of eight} & \multicolumn{3}{c}{Best of eight}\\
\cmidrule(lr){2-3}\cmidrule(lr){4-6}\cmidrule(l){7-9}
Method & Recovery $\uparrow$ & Unique $\uparrow$ & $\bothfour\uparrow$ & scRMSD $\downarrow$ & scTM $\uparrow$ & $\bothfour\uparrow$ & scRMSD $\downarrow$ & scTM $\uparrow$\\
\midrule
\grnade{} & \meanstd{41.69}{0.13} & \meanstd{6.14}{0.02} & \meanstd{7.83}{0.31} & \meanstd{16.39}{0.03} & \bestmeanstd{0.258}{0.0011} & \meanstd{11.18}{0.70} & \meanstd{13.60}{0.16} & \meanstd{0.289}{0.0029}\\
NA-MPNN & \bestmeanstd{42.65}{0.02} & \meanstd{7.22}{0.06} & \meanstd{6.68}{0.16} & \meanstd{16.68}{0.05} & \meanstd{0.248}{0.0007} & \meanstd{11.78}{0.35} & \meanstd{13.65}{0.17} & \meanstd{0.288}{0.0006}\\
\riboflowmix{} & \meanstd{32.14}{0.20} & \meanstd{6.96}{0.01} & \meanstd{4.12}{0.09} & \meanstd{18.62}{0.05} & \meanstd{0.192}{0.0011} & \meanstd{9.02}{0.35} & \meanstd{13.75}{0.24} & \meanstd{0.219}{0.0008}\\
\laribobase{} & \meanstd{34.43}{0.06} & \bestmeanstd{7.67}{0.02} & \meanstd{7.95}{0.09} & \meanstd{16.36}{0.15} & \meanstd{0.244}{0.0011} & \meanstd{14.24}{0.35} & \bestmeanstd{11.91}{0.21} & \meanstd{0.309}{0.0015}\\
\laribotri{} & \meanstd{34.82}{0.10} & \meanstd{7.62}{0.02} & \bestmeanstd{8.84}{0.13} & \bestmeanstd{15.90}{0.17} & \meanstd{0.252}{0.0004} & \bestmeanstd{14.84}{0.70} & \meanstd{12.11}{0.13} & \bestmeanstd{0.313}{0.0020}\\
\bottomrule
\end{tabular*}
\end{table}

\begin{figure}[!t]
    \centering
    \includegraphics[width=0.9\linewidth]{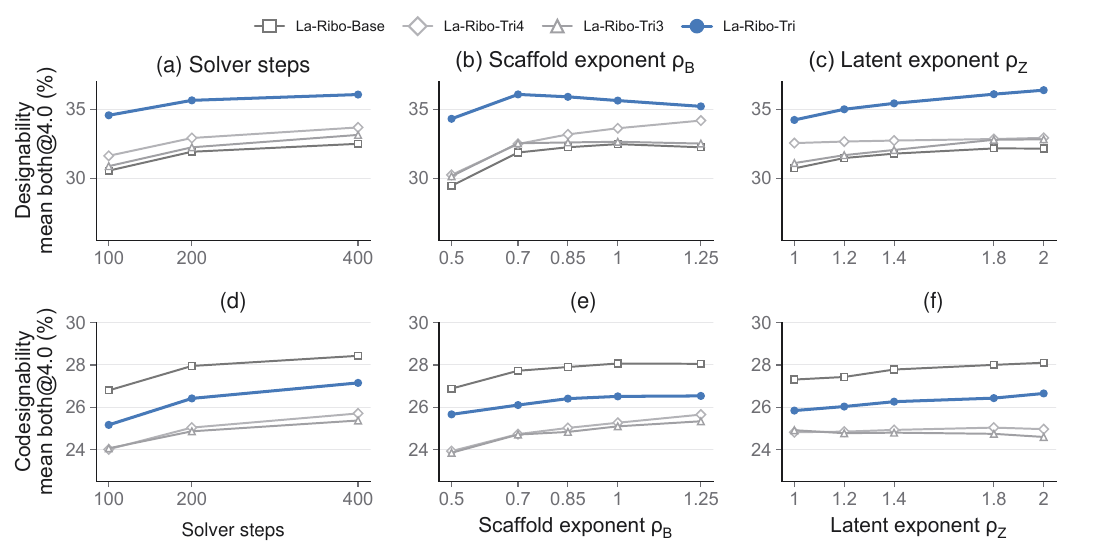}
    \caption{Sampling sensitivity. Columns show solver steps, scaffold power $\rho_B$, and latent power $\rho_Z$; top/bottom rows show marginal strict $\bothfour$ designability/codesignability.}
    \label{fig:sampling_factor_response_main}
\end{figure}

\paragraph{Sampling sensitivity}
\label{sec:ablation_sampling_sensitivity}
We vary solver steps $K$ and schedule powers $\rho_B,\rho_Z$ with 600 RNAs per setting. Averaged over the remaining factors, strict designability varies by under two points between 100 and 400 steps and changes by under 0.4 points across $\rho_B\in[0.70,1.25]$ (\cref{fig:sampling_factor_response_main}), so the reported results do not depend on a narrowly tuned schedule. Performance increases weakly with $\rho_Z$ and our default sits at the top of the swept range; \cref{app:schedule_sensitivity} shows the full surfaces.

\subsection{Scaffold-conditioned inverse folding}
\label{sec:conditional_sequence}

\textbf{Setup.} Inverse-folding targets are drawn from the validation partition, whose sequence clusters are disjoint from all \laribo{} training data (\cref{app:data_inventory}): its complete experimental subset (1,002 records spanning 107 clusters), with eight sequences per target. La-Ribo fixes $B$ and samples $Z$ without the native sequence. \textbf{Results.} Without any conditional training, the same prior directly performs scaffold-conditioned inverse folding: \laribotri{} attains the highest best-of-eight strict success ($14.84\pm0.70\%$) and scTM (0.313), and \laribobase{} the lowest scRMSD (11.91~\AA{}), versus $11.18\pm0.70\%$ for \grnade{}, $11.78\pm0.35\%$ for NA-MPNN. Recovery orders the methods differently: \grnade{} and NA-MPNN recover above 40\%, while \riboflowmix{} and both \laribo{} variants recover 32--35\%. The comparison is conservative in training data and inputs: gRNAde and NA-MPNN were trained on PDB-derived corpora that may overlap these targets and receive native structures, whereas La-Ribo is cluster-held-out and conditions on three anchors per residue.

\section{Conclusion}
\label{sec:conclusion}

La-Ribo couples a sparse RNA scaffold with residue-wise latents in a shared geometry--latent flow for RNA sequence--structure co-design; with nucleotide identity carried by the latents, the same prior also serves as a scaffold-conditioned inverse folder without further training. To ease the structural supervision bottleneck, we built a quality-controlled corpus of 168,561 RNA structures, combining experimental data with predictions from three folding models. La-Ribo improves designability and codesignability over the evaluated baselines across sampling budgets and refolding models. Future work can condition generation on function and molecular interactions, evaluate designs experimentally, extend the corpus pipeline as new structures become available, and scale to longer RNAs, toward functional RNAs of therapeutic and synthetic-biology interest.

\bibliographystyle{iclr2027_conference}
\bibliography{refs}

\clearpage
\appendix
\setupappendixrefs
\section*{Appendix contents}
\begingroup\small
\appentry{app:data_inventory}
\appsub{app:data_construction}\appsub{app:data_features}\appsub{app:data_distribution}\medskip
\appentry{app:config}
\appsub{app:optimization}\appsub{app:exact_losses}\appsub{app:flow_details}\medskip
\appentry{app:representation_analysis}
\appsub{app:vae_latent_controls}\appsub{app:representation_ablation}\medskip
\appentry{app:extended_generation}
\appsub{app:complete_benchmark}\appsub{app:triangle_frequency}\medskip
\appentry{app:structural_quality}
\appsub{app:secondary_agreement}\appsub{app:atomistic}\appsub{app:novelty}\medskip
\appentry{app:generation_landscape}
\appsub{app:schedule_sensitivity}\appsub{app:sampling_factors}\medskip
\appentry{app:evaluation_robustness}
\appsub{app:evaluation_protocols}%
\appsub{app:opendde_robustness}\appsub{app:conditional_sequence_protocol}\medskip
\appentry{app:broader_impacts}
\endgroup

\clearpage
\begin{figure}[!t]
\centering
\includegraphics[width=\textwidth]{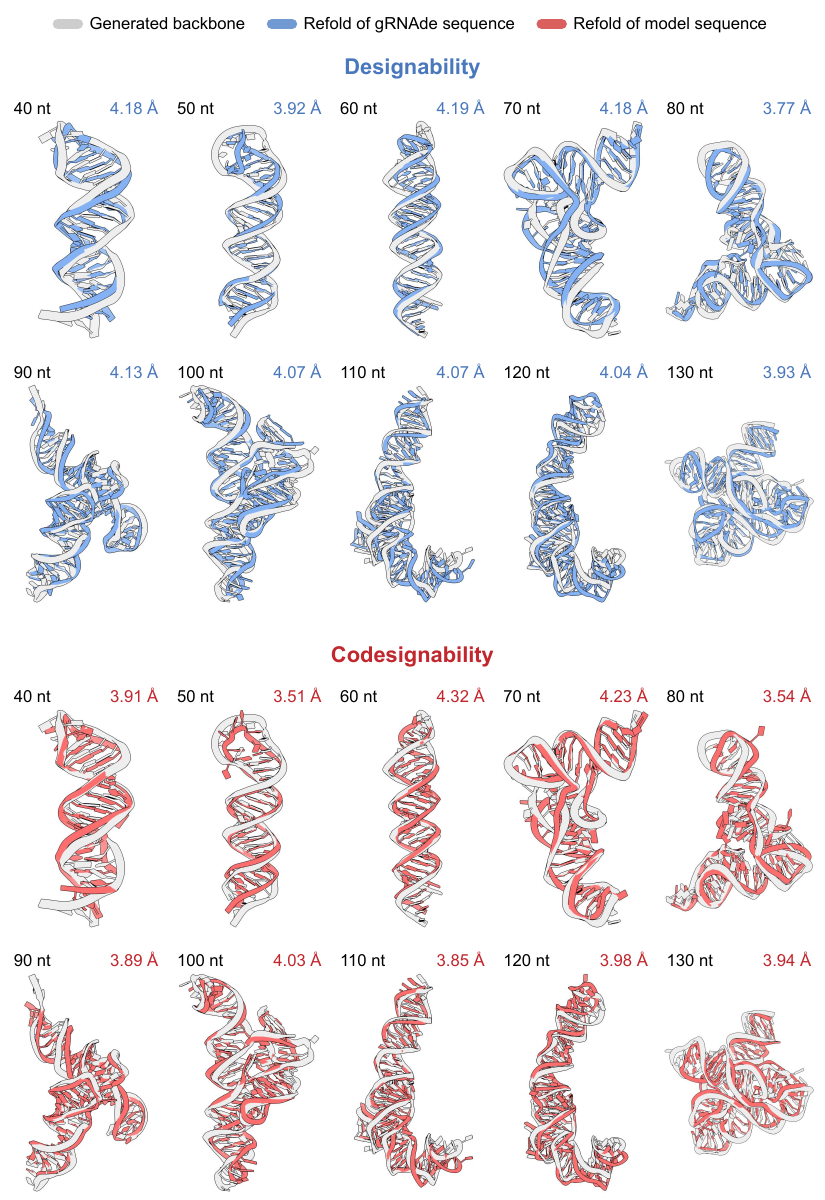}
\caption{\laribo{} backbones (40--130~nt, grey) overlaid with refolds of the best of eight \grnade{} sequences (top) and the co-designed sequence (bottom). Numbers: C4$'$ scRMSD (\AA{}).}
\label{fig:gallery}
\end{figure}

\clearpage

\section{Dataset construction and corpus statistics}
\label{app:data_inventory}

We construct the RNA corpus by combining curated experimental structures,
filtered synthetic structures from existing resources, and newly generated
trRosettaRNA2 structures. After source-specific screening and common eligibility
checks, 168,561 structures satisfy the 16--256-nucleotide length range and
minimum 90\% valid-frame coverage (\cref{tab:data_inventory}). Source-specific
preparation is summarized below; common preprocessing is detailed in
\cref{app:data_construction}.

\begin{table}[ht]
    \centering
    \caption{Eligible RNA structures by source after length and valid-frame filtering. RNASolo and the RhoFold subset are experimental; other sources are synthetic.}
    \vspace{0.4\baselineskip}
    \label{tab:data_inventory}
    \setlength{\tabcolsep}{6pt}
    \begin{tabular}{@{}llr@{}}
        \toprule
        \textbf{Source} & \textbf{Type} & \textbf{Structures} \\
        \midrule
        RNASolo & Experimental & 5,772 \\
        RhoFold subset & Experimental & 3,268 \\
        OpenFold3 Distillation & Synthetic & 125,833 \\
        NuFold Distillation & Synthetic & 23,057 \\
        trRosettaRNA2 Distillation & Synthetic & 10,631 \\
        \midrule
        \textbf{Total} & Experimental + Synthetic & \textbf{168,561} \\
        \bottomrule
    \end{tabular}
\end{table}

\paragraph{Experimental structures.}
RNASolo contributes 5,772 eligible PDB-derived RNA
chains~\citep{adamczyk2022rnasolo}. A further 3,268 experimental PDB chains
come from the RhoFold data collection~\citep{shen2024accurate}.

\paragraph{Existing synthetic resources.}
OpenFold3 Distillation uses the public RNA self-distillation
release\footnote{OpenFold3 datasets:
\url{https://portal.openfold.omsf.io/datasets}.}, comprising synthetic structures
for Rfam 15.1 cluster representatives after source confidence filtering.
The common eligibility criteria retain 125,833 structures~\citep{openfold3preview}.
NuFold Distillation uses the public NuFold Structure
Database\footnote{NuFold downloads:
\url{https://nufold.kiharalab.org/downloads}.}. We retain structures with
mean pLDDT \(\geq80\), R2DT consistency \(\geq0.70\), and MSA depth
\(\geq100\), yielding 23,057 eligible structures~\citep{kagaya2025nufold}.

\paragraph{New trRosettaRNA2 synthetic set.}
We merge sequences from bpRNA-1m(90) and
RNAStralign~\citep{danaee2018bprna,tan2017turbofold2}, restrict lengths to
16--256 nucleotides, and remove exact duplicates. We then generate
MSA-supported structures with trRosettaRNA2, retain entries with mean pLDDT
above 80, and use the relaxed top-ranked structure. This newly generated
subset contributes 10,631 eligible structures~\citep{wang2026trrosettarna2}.

\subsection{Dataset construction}
\label{app:data_construction}

The common pipeline aligns sequence and structure, extracts the geometric
representation used by La-Ribo, and defines the dataset partition.

\paragraph{Sequence--structure alignment.}
Each entry pairs a canonical sequence \(S=(s_1,\ldots,s_N)\),
\(s_i\in\mathcal A=\{A,C,G,U\}\), with a single-chain atomistic structure
\(X\in\mathbb R^{N\times37\times3}\). Residues are aligned by chain order
or, for internally unresolved positions, by residue number. Ambiguous orderings,
duplicate residue identifiers, unsupported residue names, and sequence--structure
identity mismatches are removed.

\paragraph{Geometric feature extraction.}
Resolved heavy atoms are mapped to the atom37 representation with presence
masks. For each residue, we extract the scaffold
\(b_i=[x_i^{\mathrm P};x_i^{\mathrm{C4^\prime}};x_i^{\nu_i}]\), where
\(\nu_i=\mathrm{N9}\) for purines and \(\nu_i=\mathrm{N1}\) for pyrimidines, with the canonical sugar frame
\(R_i^c=\mathcal F(x_i^{\mathrm{C4^\prime}},x_i^{\mathrm{O4^\prime}},x_i^{\mathrm{C3^\prime}})\)
and ten torsion targets. Valid-frame coverage determines structure-level
eligibility; residue, atom, and torsion masks define the entries that contribute
to autoencoder supervision.

\paragraph{Sequence-cluster partitioning.}
Sequences are clustered across all five sources with MMseqs2 at 90\% sequence
identity and 95\% coverage. Entire clusters are assigned to a single partition,
approximately 95/5, so no MMseqs2 cluster is shared across the two partitions. The resulting split contains 159,955
training and 8,606 validation records. Both scaffold-conditioned autoencoding and geometry--latent
flow training use only the training partition; the validation partition is
reserved exclusively for evaluation.

\subsection{Prepared representation}
\label{app:data_features}

\Cref{tab:prepared_features} summarizes the prepared data and geometric
quantities used by the scaffold-conditioned autoencoder. The scaffold \(B\) is
extracted from \(X\), while the latent representation \(Z\) is inferred by
\(E_\phi\). Residue, atom, and torsion masks select valid supervision targets.
Neither autoencoding nor geometry--latent flow uses MSAs or homologous structural
templates as model inputs; MSAs are used only upstream to generate part of the
synthetic corpus, while fixed nucleotide templates support local atom construction
in the decoder.

\begin{table}[ht]
    \centering
    \caption{Prepared geometry and atom features for an RNA of length \(N\); coordinates are in \AA{}.}
    \label{tab:prepared_features}
    \vspace{0.4\baselineskip}
    \small
    \setlength{\tabcolsep}{4.5pt}
    \begin{tabular}{@{}>{\raggedright\arraybackslash}p{0.25\linewidth}
                        >{\raggedright\arraybackslash}p{0.22\linewidth}
                        p{0.47\linewidth}@{}}
        \toprule
        \textbf{Quantity} & \textbf{Shape} & \textbf{Definition and role} \\
        \midrule
        Sequence \(S\)
            & \(N\)
            & Canonical nucleotide identities \(s_i\in\{A,C,G,U\}\). \\
        \addlinespace[0.25em]
        Heavy-atom coordinates \(X\)
            & \(N\times37\times3\)
            & Atom37-style coordinates; \(x_i^a\in\mathbb R^3\) denotes atom \(a\) in residue \(i\). \\
        \addlinespace[0.25em]
        Scaffold \(B=(b_i)_{i=1}^N\)
            & \(N\times9\)
            & \(b_i=[x_i^{\mathrm P};x_i^{\mathrm{C4^\prime}};x_i^{\nu_i}]\), with \(\nu_i=\mathrm{N9}\) for A/G and \(\mathrm{N1}\) for C/U. \\
        \addlinespace[0.25em]
        Latent representation \(Z\)
            & \(N\times16\)
            & Residue-wise latent inferred by \(E_\phi(S,X,B)\); it complements \(B\) with nucleotide identity and local atomistic information. \\
        \addlinespace[0.25em]
        Canonical sugar frame \(R_i^c\)
            & \(N\times3\times3\)
            & \(\mathcal F(x_i^{\mathrm{C4^\prime}},x_i^{\mathrm{O4^\prime}},x_i^{\mathrm{C3^\prime}})\), centered at \(x_i^{\mathrm{C4^\prime}}\). \\
        \addlinespace[0.25em]
        Torsion targets
            & \(N\times10\times2\)
            & Ten RNA dihedrals represented as normalized sine--cosine pairs and used to supervise \(U_i\). \\
        \addlinespace[0.25em]
        Validity masks
            & \(N\times37\), \(N\times10\), \(N\)
            & Atom, torsion, and residue masks selecting valid supervision targets. \\
        \bottomrule
    \end{tabular}
\end{table}

\subsection{Corpus distributions and sampling}
\label{app:data_distribution}

\begin{figure}[p]
    \centering
    \includegraphics[width=\linewidth]{%
        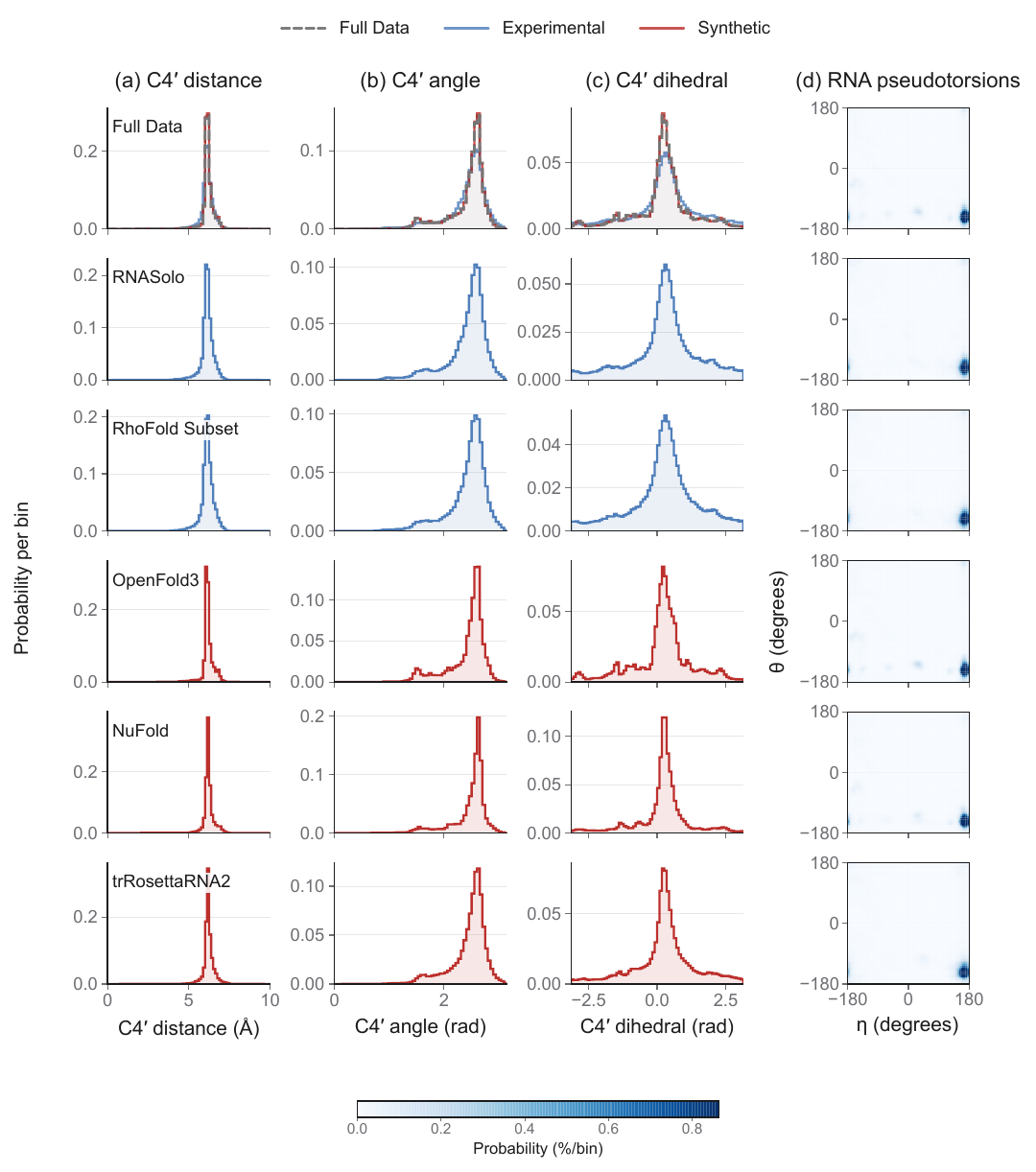}
    \caption{Local geometry by corpus source. Rows show the pooled corpus and individual sources; columns show C4$'$ distance, angle, dihedral, and joint $(\eta,\theta)$ density. In the pooled row, blue and red outlines distinguish experimental and synthetic structures. Histograms are normalized within each cohort; density panels share a color scale.}
    \label{fig:data_by_source}
\end{figure}

\begin{figure}[p]
    \centering
    \includegraphics[width=\linewidth]{%
        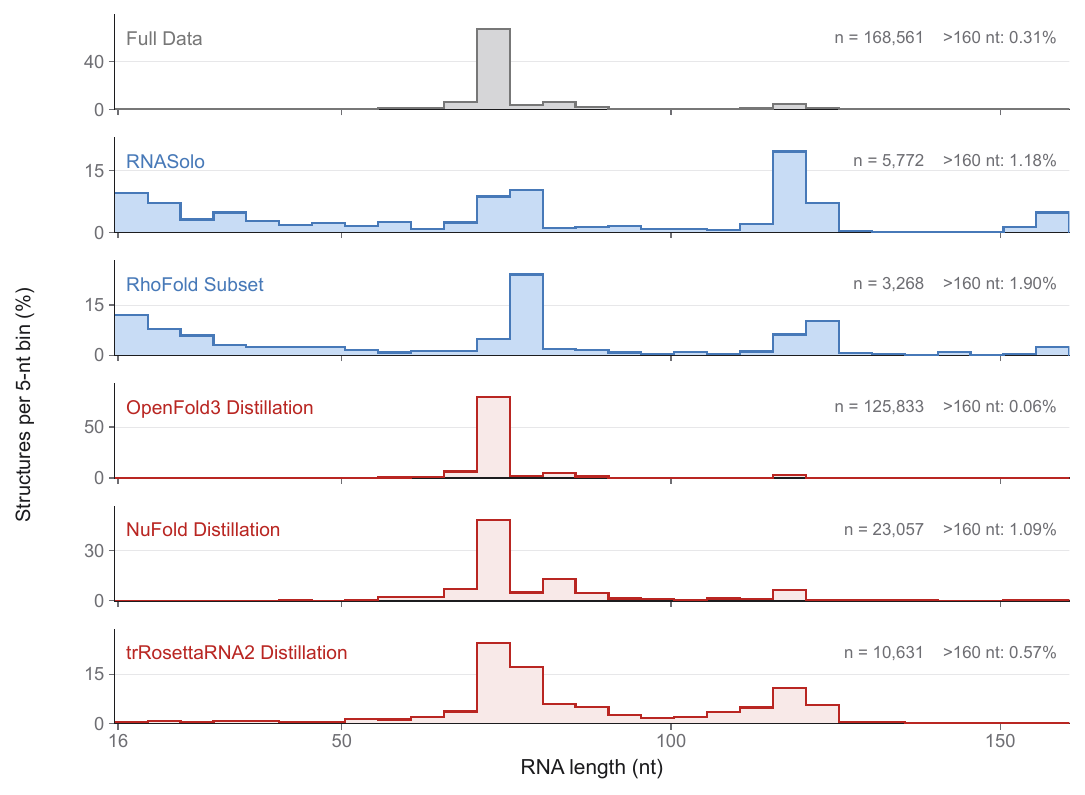}
    \caption{Length distributions of 168,561 eligible RNAs, pooled (top) and by source. Histograms are row-normalized; labels report the fraction above 160 nt. Gray, blue, and red denote pooled, experimental, and synthetic structures.}
    \label{fig:training_length_distributions}
\end{figure}

Experimental and synthetic structures share the dominant local-geometry modes
(\cref{fig:data_by_source}). One-dimensional Jensen--Shannon divergences are
below 0.055 bits, and median consecutive-\(\mathrm{C4^\prime}\) distances
differ by less than 0.014~\AA{}. The joint \((\eta,\theta)\) divergence is
0.1007 bits; among synthetic sources, OpenFold3 and trRosettaRNA2 differ most
(0.1697 bits). The length distribution peaks near 73 nucleotides because of
the large OpenFold3 collection, while the experimental, NuFold, and
trRosettaRNA2 sources provide complementary coverage
(\cref{fig:training_length_distributions}). These differences motivate
retaining both source and length information during training. Distance plots stop at 10~\AA{}, excluding fewer than 0.004\% of observations;
summary statistics retain all values. Restricting the analysis to the
training partition leaves the reported medians and circular means
unchanged.

\FloatBarrier
\section{Model and training details}
\label{app:config}

This appendix specifies the architecture, optimization, and exact training
procedures underlying the scaffold-conditioned autoencoder and geometry--latent
flow. \Cref{tab:autoencoder_hyperparameters,tab:flow_hyperparameters} summarize the stage-specific hyperparameters; subsequent sections expand the two objectives and the sampling procedure.

\subsection{Architecture and optimization}
\label{app:optimization}

\begin{table}[ht]
    \centering
    \caption{Scaffold-conditioned autoencoder hyperparameters.}
    \vspace{0.4\baselineskip}
    \label{tab:autoencoder_hyperparameters}
    \small
    \setlength{\tabcolsep}{5.2pt}
    \begin{tabular}{@{}p{0.58\linewidth}p{0.38\linewidth}@{}}
        \toprule
        Setting & Value \\
        \midrule
        Optimizer & \texttt{AdamW}, $\beta=(0.9,0.999)$ \\
        Learning rate & $10^{-4}$ \\
        Weight decay & $10^{-2}$ \\
        Epochs & 100 \\
        Batch size & 24 max (length-dependent) \\
        Gradient clipping & 1.0 \\
        Objective weights & $\lambda_{\mathrm{gt}}=0.25$, $\lambda_{\mathrm{pred}}=0.10$, $\lambda_{\mathrm{KL}}=10^{-2}$ \\
        \bottomrule
    \end{tabular}
\end{table}
\begin{table}[ht]
    \centering
    \caption{Geometry--latent flow hyperparameters and default sampling settings.}
    \vspace{0.4\baselineskip}
    \label{tab:flow_hyperparameters}
    \small
    \setlength{\tabcolsep}{5.2pt}
    \begin{tabular}{@{}p{0.58\linewidth}p{0.38\linewidth}@{}}
        \toprule
        Setting & Value \\
        \midrule
        Optimizer & \texttt{AdamW}, $\beta=(0.9,0.999)$ \\
        Learning rate & $10^{-4}$ \\
        Weight decay & $10^{-2}$ \\
        Epochs & 100 \\
        Batch size & 24 max (length-dependent) \\
        Gradient clipping & 1.0 \\
        Self-conditioning probability & 0.5 \\
        EMA decay & 0.999 \\
        Default sampling parameters & $\rho_B=1.25$, $\rho_Z=2.0$ \\
        \bottomrule
    \end{tabular}
\end{table}

\paragraph{Architecture.}
The encoder $E_\phi$ and decoder $D_\psi$ each use six token--pair blocks,
token width 256, pair width 128, and eight attention heads. The latent width
is 16, and the autoencoder contains 14.91M parameters. The flow network uses
12 blocks, token width 512, pair width 256, and eight heads. Its residue
features combine noisy scaffold coordinates, latent states, and modality
times. Pair features encode relative sequence separation and distances among
the scaffold anchors $x_i^{\mathrm P}$, $x_i^{\mathrm{C4^\prime}}$, and
$x_i^{\nu_i}$ from the noisy and self-conditioned scaffold states; these
features enter residue attention as pair-dependent biases.

\begin{figure}[t]
    \centering
    \includegraphics[width=0.95\linewidth]{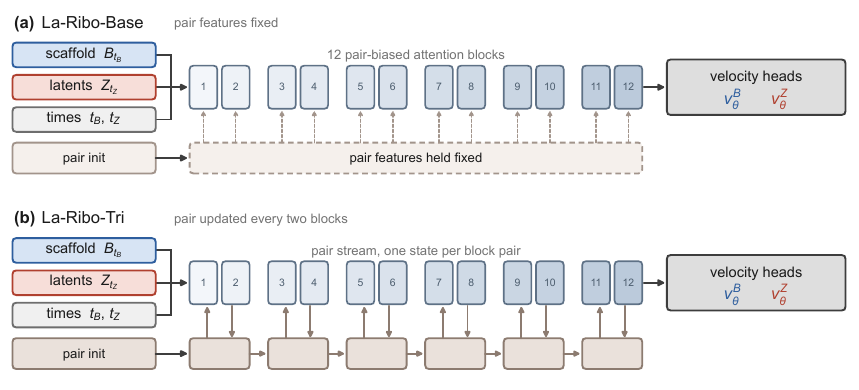}
    \caption{Base and Tri flow architectures. Both use a 12-block trunk to predict scaffold and latent velocities. \laribobase{} keeps pair features fixed as attention biases; \laribotri{} updates them every two blocks using token-to-pair projections and triangle multiplicative updates.}
    \label{fig:base_tri_architecture}
\end{figure}

\laribobase{} keeps the pair representation fixed throughout the flow trunk.
\laribotri{} additionally refines the pair stream from evolving residue
features using token-to-pair projections and triangle multiplicative updates
every two blocks. \laribotrithree{} and \laribotrifour{} use the same update
with intervals of three and four blocks. The trainable parameter counts are
58.82M for \laribobase{}, 61.46M for \laribotrifour{}, 62.77M for
\laribotrithree{}, and 65.41M for \laribotri{}. These variants therefore
compare different pair-update architectures rather than parameter-matched
models. \Cref{fig:base_tri_architecture} gives a compact architectural sketch focused on the Base--Tri difference.

\paragraph{Length-aware batching.}
To control memory growth with sequence length, training uses length buckets
with a length-dependent per-device batch cap of at most 24 sequences. Within
each bucket, source and sequence cluster are sampled hierarchically before a
cluster member is drawn.

\subsection{Scaffold-conditioned autoencoding: exact objective}
\label{app:exact_losses}

This section expands the scaffold-conditioned autoencoding objective in
\cref{sec:stage1}. We retain the main-text convention that $Q_i$, $U_i$, and
$R_i^{\mathrm{dec}}$ are decoder outputs; a superscript $\mathrm{ref}$
identifies a target torsion where a distinction is needed.

\paragraph{Local frames.}
For noncollinear anchors, the deterministic operator
\(\mathcal F(o,u,v)=[e_1,e_2,e_3]\) constructs a right-handed orientation:
\begin{equation}
\begin{aligned}
e_1&=\frac{u-o}{\|u-o\|_2},&
e_2&=\frac{(v-o)-e_1e_1^\top(v-o)}
{\|(v-o)-e_1e_1^\top(v-o)\|_2},&
e_3&=e_1\times e_2.
\end{aligned}
\label{eq:exact_frame_operator}
\end{equation}
Basis vectors are matrix columns; the origin \(o\) is stored separately.
The scaffold pseudo-frame is built from the input scaffold and the target
sugar frame from atomistic coordinates:
\begin{equation}
\begin{aligned}
R_i^p&=\mathcal F(x_i^{\mathrm{C4^\prime}},
x_i^{\nu_i},x_i^{\mathrm{P}}),&
R_i^c&=\mathcal F(x_i^{\mathrm{C4^\prime}},
x_i^{\mathrm{O4^\prime}},x_i^{\mathrm{C3^\prime}}).
\end{aligned}
\label{eq:exact_frames}
\end{equation}
Both frames use the same \(\mathrm{C4^\prime}\) origin. The target relative
rotation is \((R_i^p)^\top R_i^c\). The decoder's six-dimensional rotation output~\citep{zhou2019continuity} is orthonormalized into \(Q_i\in\mathrm{SO}(3)\), giving
\(R_i^{\mathrm{dec}}=R_i^pQ_i\). Degenerate inputs use the implementation's
deterministic numerical guards. The three scaffold anchors need not form a
rigid chemical group to define this coordinate frame.

\begin{table}[ht]
    \centering
    \caption{Atom-construction torsions and their standard relations. Each quartet defines a reconstruction dihedral; ``reversed'' reverses its order without changing the angle, and ``shifted'' replaces terminal atoms. Here $\nu_i$ follows \cref{eq:scaffold}, and $A_{\mathrm{base}}$ is $\mathrm{C4}$ for A/G or $\mathrm{C2}$ for C/U.}
    \vspace{0.4\baselineskip}
    \label{tab:rna_torsions}
    \footnotesize
    \begin{tabular}{cll}
        \toprule
        Index & Atom quartet & Standard relation \\
        \midrule
        1 & \(\mathrm{C5^\prime}-\mathrm{C4^\prime}-\mathrm{C3^\prime}-\mathrm{C2^\prime}\) & \(\delta\) shifted (\(\mathrm{C2^\prime}\) for \(\mathrm{O3^\prime}\)) \\
        2 & \(\mathrm{C3^\prime}-\mathrm{C4^\prime}-\mathrm{O4^\prime}-\mathrm{C1^\prime}\) & \(\nu_4\) \\
        3 & \(\mathrm{C4^\prime}-\mathrm{O4^\prime}-\mathrm{C1^\prime}-\nu_i\) & \(\nu_0\) shifted (\(\nu_i\) for \(\mathrm{C2^\prime}\)) \\
        4 & \(\mathrm{C5^\prime}-\mathrm{C4^\prime}-\mathrm{C3^\prime}-\mathrm{O3^\prime}\) & \(\delta\) \\
        5 & \(\mathrm{C3^\prime}-\mathrm{C4^\prime}-\mathrm{C5^\prime}-\mathrm{O5^\prime}\) & \(\gamma\) reversed \\
        6 & \(\mathrm{C4^\prime}-\mathrm{C5^\prime}-\mathrm{O5^\prime}-\mathrm P\) & \(\beta\) reversed \\
        7 & \(\mathrm{C5^\prime}-\mathrm{O5^\prime}-\mathrm P-\mathrm{OP1}\) & \(\alpha\) reversed and shifted (\(\mathrm{OP1}\) for \(\mathrm{O3^\prime}_{i-1}\)) \\
        8 & \(\mathrm{C5^\prime}-\mathrm{O5^\prime}-\mathrm P-\mathrm{OP2}\) & \(\alpha\) reversed and shifted (\(\mathrm{OP2}\) for \(\mathrm{O3^\prime}_{i-1}\)) \\
        9 & \(\nu_i-\mathrm{C1^\prime}-\mathrm{C2^\prime}-\mathrm{O2^\prime}\) & \(\nu_1\) shifted (\(\nu_i\), \(\mathrm{O2^\prime}\)) \\
        10 & \(\mathrm{C2^\prime}-\mathrm{C1^\prime}-\nu_i-A_{\mathrm{base}}\) & \(\chi\) shifted (\(\mathrm{C2^\prime}\) for \(\mathrm{O4^\prime}\)) \\
        \bottomrule
    \end{tabular}
\end{table}

\paragraph{Torsion-based atom construction.}
\Cref{tab:rna_torsions} specifies the ten dihedrals used in extraction and
reconstruction. Entries 1 and 4 share the \(\mathrm{C4^\prime}-\mathrm{C3^\prime}\) bond, and entries
7 and 8 share \(\mathrm{O5^\prime}-\mathrm P\); the ten entries therefore span eight bonds and allow
separate placement of substituent branches. Inter-residue \(\epsilon\)
and \(\zeta\) are not construction variables because each residue is built
from its own sugar frame. For entries 9 and 10, upstream sugar rotations
alter the atoms defining the local frame. We rebuild their zero-angle frames from the current \(\mathrm{C1^\prime}\), \(\mathrm{C2^\prime}\), and \(\nu_i\) coordinates.

\paragraph{Template expansion and gradient paths.}
One decoder forward pass produces nucleotide probabilities \(\pi\), relative
rotations \(Q\), and normalized sine--cosine torsion vectors \(U\).
The same frames and torsions are expanded twice: the \(\mathrm{gt}\) view
uses native identities \(s_i\), and the \(\mathrm{pred}\) view uses
\(s_i^{\mathrm{pred}}=\arg\max_{a\in\mathcal A}\pi_i(a)\).
The \(\mathrm{C4^\prime}\) coordinate is copied as the frame origin; the \(\mathrm P\) and \(\nu_i\) anchors are reconstructed.
Both views are supervised against the reference structure.

Predicted-template selection is non-differentiable. Its geometric losses
update the encoder, shared decoder, and geometric output heads, but do not
backpropagate through argmax to the nucleotide logits. The nucleotide head
is supervised by cross-entropy. Sequence, rotation, and torsion losses are
included once; the second template view adds only geometric reconstruction
terms.

\paragraph{Template-independent losses.}
Let \(\mathcal R\) be the valid residue set and \(\mathcal T\) the valid
residue--torsion pairs. For target angles \(\tau_{ik}\), define
\(u_{ik}^{\mathrm{ref}}=(\sin\tau_{ik},\cos\tau_{ik})\); \(u_{ik}\)
is the corresponding normalized decoder output. The losses are
\begin{equation}
\begin{aligned}
\mathcal L_{\mathrm{seq}}
&=-\frac{1}{|\mathcal R|}\sum_{i\in\mathcal R}\log\pi_i(s_i),\\
\mathcal L_{\mathrm{rot}}
&=\frac{1}{|\mathcal R|}\sum_{i\in\mathcal R}
d_{\mathrm{SO}(3)}^2(R_i^{\mathrm{dec}},R_i^c),\\
\mathcal L_{\mathrm{tors}}
&=\frac{1}{2|\mathcal T|}\sum_{(i,k)\in\mathcal T}
\|u_{ik}-u_{ik}^{\mathrm{ref}}\|_2^2.
\end{aligned}
\label{eq:exact_shared_losses}
\end{equation}
Here \(d_{\mathrm{SO}(3)}\) is the geodesic rotation angle. All sums exclude
masked entries. The factor \(1/2\) in the torsion loss averages the two
sine--cosine components.

\paragraph{Template-dependent geometric losses.}
For \(v\in\{\mathrm{gt},\mathrm{pred}\}\), let \(M_i\) be the valid target
atom names and \(M_i^v\) those instantiated by template \(v\).
Write \(x_i^{a,v}\) for a reconstructed atom and \(b_i^v\) for the
scaffold extracted from that reconstruction using the anchors in \cref{eq:scaffold}.
The glycosidic anchor is selected according to the instantiated nucleotide,
so it remains an attachment-site comparison even when base type changes.
We use
\begin{equation}
\begin{aligned}
\mathcal L_{\mathrm{atom}}^v
&=\frac{\sum_{i\in\mathcal R}\sum_{a\in M_i\cap M_i^v}
\|x_i^{a,v}-x_i^a\|_2^2}{\sum_{i\in\mathcal R}|M_i\cap M_i^v|},\\
\mathcal L_{\mathrm{scaf}}^v
&=\frac{1}{9|\mathcal R|}\sum_{i\in\mathcal R}\|b_i^v-b_i\|_2^2,\\
\mathcal L_{\mathrm{pseudo}}^v
&=\frac12\left[
\ell_{\mathrm{circ}}(\eta^v,\eta;\Omega_\eta)
+\ell_{\mathrm{circ}}(\theta^v,\theta;\Omega_\theta)\right].
\end{aligned}
\label{eq:exact_view_losses}
\end{equation}
The atom term averages squared Euclidean distances over shared atom names;
unmatched atoms are not assigned zero-coordinate targets. The scaffold
term averages its nine coordinate components against the reference scaffold.
For a periodic angle, the circular loss is
\begin{equation}
\ell_{\mathrm{circ}}(\alpha^v,\alpha;\Omega)
=\frac{1}{|\Omega|}\sum_{j\in\Omega}
\|c(\alpha_j^v)-c(\alpha_j)\|_2^2,
\qquad c(\alpha)=(\sin\alpha,\cos\alpha).
\label{eq:circular_loss}
\end{equation}
\(\Omega_\eta\) and \(\Omega_\theta\) contain the measurable trace
pseudotorsions. Using coordinate arguments throughout,
\begin{equation}
\begin{aligned}
\eta_i&=\operatorname{dih}(x_{i-1}^{\mathrm{C4^\prime}},x_i^{\mathrm P},
x_i^{\mathrm{C4^\prime}},x_{i+1}^{\mathrm P}),\\
\theta_i&=\operatorname{dih}(x_i^{\mathrm P},x_i^{\mathrm{C4^\prime}},
x_{i+1}^{\mathrm P},x_{i+1}^{\mathrm{C4^\prime}}).
\end{aligned}
\label{eq:pseudotorsions}
\end{equation}
These follow the RNA pseudotorsional convention of
\citet{duarte1998stepping,wadley2007evaluating}. Reconstructed values use
\(X^v\); joint distributions pair \(\eta\) and \(\theta\) at the same residue.
Their three geometric losses sum to \(\mathcal L_{\mathrm{geom}}^v\)
in \cref{eq:geometry_ae_loss}.

\paragraph{KL regularization and weights.}
The residue-wise Gaussian parameters from \(E_\phi\) are regularized by
\begin{equation}
\mathcal L_{\mathrm{KL}}
=\frac{1}{2|\mathcal R|}\sum_{i\in\mathcal R}\sum_{d=1}^{16}
\left(\mu_{id}^2+\sigma_{id}^2-\log\sigma_{id}^2-1\right).
\label{eq:exact_kl}
\end{equation}
Latent dimensions are summed within each residue and then averaged over
valid residues. The small KL coefficient trades off latent regularization and
reconstruction accuracy.

\subsection{Geometry--latent flow: training and sampling}
\label{app:flow_details}

The geometry--latent flow is trained after freezing $E_\phi$ and $D_\psi$.
\Cref{alg:flow_training,alg:flow_sampling} summarize training and unconditional sampling
procedures using the same states and updates as the main-text formulation.
Exact normalization, scaffold recentering, time sampling, and endpoint
stabilization are specified immediately afterward; $\operatorname{sg}$ denotes stop-gradient.

\begin{algorithm}[H]
\caption{\laribo~Training}
\label{alg:flow_training}
\begin{algorithmic}[1]
\setlength{\itemsep}{4pt}
\Require RNA example $(S,X,B)$; frozen encoder $E_\phi$; velocity network $v_\theta$
\Ensure Updated flow parameters $\theta$
\State $\mu \gets E_\phi(S,X,B)$
\State $(x_1^B,x_1^Z) \gets \operatorname{Normalize}(B,\mu)$
\State $\epsilon^B,\epsilon^Z\sim\mathcal N(0,I)$; $x_0^B\gets\Pi_B\epsilon^B$; $x_0^Z\gets\epsilon^Z$
\State Sample $t_B,t_Z\sim p_t$ independently
\For{$a\in\{B,Z\}$}
    \State $x_{t_a}^a\gets(1-t_a)x_0^a+t_a x_1^a$ \Comment{Eq.~\ref{eq:geometry_latent_path}}
\EndFor
\State $x_{\mathrm{sc}}\gets 0$
\State $u\sim\operatorname{Bernoulli}(0.5)$
\If{$u=1$}
    \State $(\widetilde v^B,\widetilde v^Z)\gets v_\theta(x_t,0,t_B,t_Z)$
    \For{$a\in\{B,Z\}$}
        \State $x_{\mathrm{sc}}^a\gets\operatorname{sg}\!\left[x_{t_a}^a+(1-t_a)\widetilde v^a\right]$
    \EndFor
\EndIf
\State $(v^B,v^Z)\gets v_\theta(x_t,x_{\mathrm{sc}},t_B,t_Z)$
\For{$a\in\{B,Z\}$}
    \State $x_{1,\theta}^a\gets x_{t_a}^a+(1-t_a)v^a$ \Comment{Eq.~\ref{eq:flow_endpoint}}
\EndFor
\State $\displaystyle \mathcal L_{\mathrm{FM}}\gets
\sum_{a\in\{B,Z\}}
\frac{\sum_{i\in\mathcal R}\|(x_{1,\theta}^a-x_1^a)_i\|_2^2}
{d_a|\mathcal R|\bigl((1-t_a)^2+\varepsilon_{\mathrm{stab}}\bigr)}$
\State Update $\theta\gets\operatorname{AdamW}(\theta,\nabla_\theta\mathcal L_{\mathrm{FM}})$
\end{algorithmic}
\end{algorithm}

\begin{algorithm}[H]
\caption{\laribo~Unconditional Sampling}
\label{alg:flow_sampling}
\begin{algorithmic}[1]
\setlength{\itemsep}{2pt}
\Require Length $N$; steps $K$; powers $(\rho_B,\rho_Z)$; trained $v_\theta$; frozen decoder $D_\psi$
\Ensure Generated nucleotide sequence $\widehat S$ and atomistic coordinates $\widehat X$
\State $\epsilon^B,\epsilon^Z\sim\mathcal N(0,I)$; $x_0^B\gets\Pi_B\epsilon^B$; $x_0^Z\gets\epsilon^Z$
\State $x_{\mathrm{sc}}\gets 0$
\For{$k=0,\ldots,K-1$}
    \State $t_k^Z\gets(k/K)^{\rho_Z}$; $t_{k+1}^Z\gets((k+1)/K)^{\rho_Z}$
    \State $t_k^B\gets(k/K)^{\rho_B}$; $t_{k+1}^B\gets((k+1)/K)^{\rho_B}$ \Comment{Eq.~\ref{eq:sampling_schedule}}
    \State $(v_k^B,v_k^Z)\gets v_\theta(x_k,x_{\mathrm{sc}},t_k^B,t_k^Z)$
    \For{$a\in\{B,Z\}$}
        \State $x_{1,k}^a\gets x_{t_k^a}^a+(1-t_k^a)v_k^a$
    \EndFor
    \State $x_{t_{k+1}^B}^B\gets x_{t_k^B}^B+(t_{k+1}^B-t_k^B)v_k^B$
    \State $x_{t_{k+1}^Z}^Z\gets x_{t_k^Z}^Z+(t_{k+1}^Z-t_k^Z)v_k^Z$ \Comment{Eq.~\ref{eq:euler_update}}
    \State $x_{\mathrm{sc}}\gets\operatorname{sg}(x_{1,k})$
\EndFor
\State $(B,Z)\gets\operatorname{Denormalize}(x_1^B,x_1^Z)$
\State $(\pi,Q,U)\gets D_\psi(B,Z)$; $\widehat S_i\gets\arg\max_{a\in\mathcal A}\pi_i(a)$
\State $\widehat X\gets\operatorname{AtomConstruct}(B,\widehat S,Q,U)$
\State \Return $(\widehat S,\widehat X)$
\end{algorithmic}
\end{algorithm}

\paragraph{Flow normalization and recentering.}
The flow targets are posterior means and scaffold coordinates.
Each latent dimension is standardized by its empirical mean and standard
deviation; the \(\mathrm{C4^\prime}\)-centered scaffold is divided by a single global RMS
scale. Let \(Y\) be any scaffold-shaped coordinate or velocity array. Define
\begin{equation}
[\Pi_B Y]_i^{a}=Y_i^{a}
-\frac{1}{|\mathcal R|}\sum_{j\in\mathcal R}Y_j^{\mathrm{C4^\prime}},
\quad a\in\{\mathrm P,\mathrm{C4^\prime},\nu_i\},
\qquad \Pi_Z=I.
\label{eq:scaffold_centering}
\end{equation}
This removes a common translation from all three anchors. Independent
standard Gaussian arrays are used for the two base states, with
\(x_0^a=\Pi_a\epsilon^a\) for modality \(a\in\{B,Z\}\).
Thus the scaffold base is a centered Gaussian, not an unconstrained
standard normal in the full coordinate space.

The main-text velocity \(v_\theta\) includes this recentering. If
\(f_\theta\) denotes the raw network outputs, then
\begin{equation}
v_\theta^a(x_t,t_B,t_Z)
=\Pi_a f_\theta^a(x_t,x_{\mathrm{sc}},t_B,t_Z).
\label{eq:effective_velocity}
\end{equation}
Here \(x_{\mathrm{sc}}\) is a detached endpoint estimate supplied by
self-conditioning, omitted from the main-text argument list. It is used
with probability 0.5 during training. For each modality independently,
time is drawn from an equal mixture of uniform and \(\operatorname{Beta}(2,2)\)
distributions on \([0,1]\), then rescaled to \([0.01,0.99]\).

\paragraph{Stabilized endpoint objective.}
Define the predicted endpoint, with \(\varepsilon_{\mathrm{stab}}=10^{-5}\), as
\begin{equation}
x_{1,\theta}^a=x_{t_a}^a+(1-t_a)v_\theta^a(x_t,t_B,t_Z).
\label{eq:flow_endpoint}
\end{equation}
The implemented objective is
\begin{equation}
\sum_{a\in\{B,Z\}}\mathbb E\!\left[
\frac{\sum_{i\in\mathcal R}\|(x_{1,\theta}^a-x_1^a)_i\|_2^2}
{d_a|\mathcal R|\bigl((1-t_a)^2+\varepsilon_{\mathrm{stab}}\bigr)}
\right].
\label{eq:stabilized_flow_loss}
\end{equation}
Without the stabilizer this equals \cref{eq:flow_loss}; the implemented
stabilizer multiplies each squared velocity error by
\((1-t_a)^2/((1-t_a)^2+\varepsilon_{\mathrm{stab}})\).
No decoded atom or sequence loss is added during flow training.

\paragraph{Euler updates and conditional sampling.}
At step \(k\), one shared network evaluation returns both velocities.
Each modality uses its own step size,
\begin{equation}
x_{t_{k+1}^a}^a=x_{t_k^a}^a+
(t_{k+1}^a-t_k^a)v_\theta^a(x_k,t_k^B,t_k^Z),
\qquad
x_k=(x_{t_k^B}^B,x_{t_k^Z}^Z).
\label{eq:euler_update}
\end{equation}
The effective scaffold velocity enforces recentering, and the previous
endpoint estimate supplies self-conditioning. The \(K\) Euler steps
therefore require \(K\) shared velocity-network evaluations.
Before decoding, coordinate scaling and latent standardization are reversed.
For scaffold-conditioned inverse folding, the observed scaffold is fixed at $t_B=1$ and only the latent modality is integrated. The resulting latent state is decoded together with the supplied scaffold, providing an approximate conditional sampler without a separately trained model.

\section{Additional representation analyses}
\label{app:representation_analysis}

\subsection{Latent interventions}
\label{app:vae_latent_controls}

The intervention conditions are paired by structure on the
cluster-held-out validation set. Scaffold $B$ stays fixed. For atom-level
reconstruction we use each RNA's reference nucleotide identity in every
condition, whereas sequence recovery is computed from the decoder's
nucleotide probabilities. The within-chain shuffle is separate from the
single-residue locality perturbation.
Non-scaffold RMSD excludes $x_i^{\mathrm P}$,
$x_i^{\mathrm{C4^\prime}}$, and $x_i^{\nu_i}$. The $x_i^{\mathrm{C4^\prime}}$
anchor is omitted from the anchor-error breakdown because it is copied as
the frame origin.
Under both regimes, zeroing or shuffling lowers sequence recovery and
increases reconstruction error, demonstrating that latent values and their
residue assignments contribute information beyond the supplied scaffold.
\Cref{fig:vae_latent_interventions} visualizes the VAE interventions;
\cref{tab:vae_latent_interventions} reports the expanded metrics for both models.

\begin{table}[!htbp]
    \centering
    \caption{Reconstruction under latent interventions.}
    \vspace{0.4\baselineskip}
    \label{tab:vae_latent_interventions}
    \footnotesize
    \setlength{\tabcolsep}{2.2pt}
    \begin{tabular}{llrrrrr}
        \toprule
        Model & Latent
        & \shortstack{Sequence\\(\%) $\uparrow$}
        & \shortstack{Frame\\($^\circ$) $\downarrow$}
        & \shortstack{Non-scaffold\\RMSD (\AA{}) $\downarrow$}
        & P (\AA) \(\downarrow\)
        & \shortstack{Glycosidic N\\(\AA{}) $\downarrow$} \\
        \midrule
        VAE & posterior
            & \meanstd{100.00}{0.00} & \meanstd{1.779}{0.010}
            & \meanstd{0.08746}{0.00021} & \meanstd{0.08111}{0.00023}
            & \meanstd{0.04709}{0.00048} \\
        VAE & zero
            & \meanstd{24.77}{0.73} & \meanstd{16.447}{0.081}
            & \meanstd{0.77254}{0.00651} & \meanstd{0.64596}{0.02262}
            & \meanstd{0.12169}{0.02587} \\
        VAE & shuffle
            & \meanstd{27.54}{0.02} & \meanstd{20.699}{0.009}
            & \meanstd{0.96503}{0.00650} & \meanstd{0.75054}{0.00275}
            & \meanstd{0.10439}{0.00108} \\
        \midrule
        E-VAE & posterior
            & \meanstd{100.00}{0.00} & \meanstd{2.673}{0.000}
            & \meanstd{0.12248}{0.00000} & \meanstd{0.09940}{0.00000}
            & \meanstd{0.05380}{0.00000} \\
        E-VAE & zero
            & \meanstd{24.99}{0.00} & \meanstd{16.910}{0.000}
            & \meanstd{0.87019}{0.00000} & \meanstd{0.69486}{0.00000}
            & \meanstd{0.11057}{0.00000} \\
        E-VAE & shuffle
            & \meanstd{27.58}{0.21} & \meanstd{20.669}{0.052}
            & \meanstd{0.98181}{0.00219} & \meanstd{0.76289}{0.00061}
            & \meanstd{0.11122}{0.00024} \\
        \bottomrule
    \end{tabular}
\end{table}

To test locality, we perturb one valid residue's latent while keeping
the scaffold and all other latents fixed. We add
\(\Delta z=m(s_z\odot d)\) at that residue: \(m\) controls strength,
\(s_z\) is the autoencoder-specific empirical standard deviation of
posterior-mean latent coordinates, and \(d\) is a unit-RMS Gaussian direction
held fixed across magnitudes. We compare non-scaffold atom displacement at
the edited residue with displacement elsewhere. The much smaller response
outside the edited residue indicates a predominantly local effect
(\cref{fig:vae_latent_interventions}c).

\subsection{Scaffold choice and latent dimensionality}
\label{app:representation_ablation}

We compare scaffold and latent-width variants on the same validation dataset. All variants are trained on experimental structures and evaluated by posterior-mean decoding.

Atoms are expanded with ground-truth nucleotide templates. For every
representation, non-scaffold RMSD excludes P, C4$'$, and the glycosidic N.
The C4$'$-only variant constructs its frame from the chain tangent and the
plane of adjacent C4$'$ steps. The P/C4$'$ variant uses the C4$'$-to-P
direction as its first axis and the orthogonalized chain tangent as its
second; both use C4$'$ as the origin.

\begin{table}[!htbp]
    \centering
    \caption{Reconstruction by scaffold and latent width $d_Z$ using experimental-only training and a common validation subset. Scaffold labels $\mathrm{P}$, $\mathrm{C4^\prime}$, and $\nu_i$ abbreviate the anchors $x_i^{\mathrm P}$, $x_i^{\mathrm{C4^\prime}}$, and $x_i^{\nu_i}$. AA/NS denote all-atom/non-scaffold RMSD (unaligned mean/median, \AA{}); Frame is angular error and Torsion is MAE (degrees, averaged over valid residues); Seq is sequence recovery in percent. Bold marks the best geometric value per column.}
    \vspace{0.4\baselineskip}
    \label{tab:representation_reconstruction}
    \small
    \setlength{\tabcolsep}{2pt}
    \begin{tabular*}{\linewidth}{@{\extracolsep{\fill}}lcrrrrr@{}}
        \toprule
        Scaffold & $d_Z$
        & AA-RMSD $\downarrow$
        & NS-RMSD $\downarrow$
        & Frame $\downarrow$
        & Torsion $\downarrow$
        & Seq $\uparrow$ \\
        \midrule
        $\mathrm{C4^\prime}$ & 16
            & 0.1533 / 0.1444 & 0.1589 / 0.1498 & 2.389 & 2.326 & 99.999 \\
        $\mathrm P,\mathrm{C4^\prime}$ & 16
            & 0.1234 / 0.1173 & 0.1304 / 0.1240 & \textbf{2.313} & \textbf{2.132} & 100.000 \\
        $\mathrm P,\mathrm{C4^\prime},\nu_i$ & 8
            & 0.1259 / 0.1198 & 0.1334 / 0.1267 & 2.516 & 2.418 & 100.000 \\
        $\mathrm P,\mathrm{C4^\prime},\nu_i$ & 16
        & \textbf{0.1159 / 0.1107} & \textbf{0.1225 / 0.1170}
        & 2.673 & 2.649 & 100.000 \\
        $\mathrm P,\mathrm{C4^\prime},\nu_i$ & 32
        & 0.1267 / 0.1117 & 0.1227 / 0.1176 & 2.886 & 2.891 & 100.000 \\
        \bottomrule
    \end{tabular*}
\end{table}

The three-anchor scaffold reduces all-atom and non-scaffold RMSD
relative to the one- and two-anchor alternatives
(\cref{tab:representation_reconstruction}). Among the three-anchor variants,
width 16 gives the lowest RMSD values (mean non-scaffold RMSD 0.1225~\AA{},
versus 0.1334 and 0.1227~\AA{} at widths 8 and 32); all generation experiments
use the three-anchor scaffold with width 16.

\section{Extended generation results and controlled ablations}
\label{app:extended_generation}
\subsection{Complete benchmark metrics}
\label{app:complete_benchmark}
The following tables expand the common protocol in \cref{sec:data,tab:main}. Success rates are percentages; continuous errors and cluster counts are reported alongside thresholded endpoints. Sampling steps are velocity-network evaluations for the forward-Euler solver.
\begin{table}[!htbp]
    \centering
    \caption{Expanded unconditional-generation metrics under the protocol of \cref{tab:main}. Rates are percentages and scRMSD is in \AA{}; bold marks the best mean across operating points.}
    \vspace{0.4\baselineskip}
    \label{tab:main_full}
    \footnotesize
    \setlength{\tabcolsep}{2.5pt}
    \textbf{(a) Designability}\\[0.45em]
    \begin{tabular}{lrrrrrrr}
        \toprule
        Method & Steps & \(\bothfour\,\uparrow\) & \(\bothfive\,\uparrow\)
        & \(\tmfortyfive\,\uparrow\) & scRMSD \(\downarrow\)
        & scTM \(\uparrow\) & \#Clust. \(\uparrow\) \\
        \midrule
        \multirow{4}{*}{\rnff{}}
            & 50 & \meanstd{25.17}{1.09} & \meanstd{29.89}{0.42}
            & \meanstd{36.61}{0.19} & \meanstd{11.58}{0.47}
            & \meanstd{0.406}{0.006} & \meanstd{24.00}{1.73} \\
            & 100 & \meanstd{26.61}{1.18} & \meanstd{32.17}{1.48}
            & \meanstd{39.67}{0.33} & \meanstd{11.36}{0.16}
            & \meanstd{0.417}{0.003} & \meanstd{27.00}{2.65} \\
            & 200 & \meanstd{26.17}{1.76} & \meanstd{31.56}{2.10}
            & \meanstd{38.44}{1.84} & \meanstd{11.47}{0.29}
            & \meanstd{0.413}{0.008} & \meanstd{26.00}{1.00} \\
            & 400 & \meanstd{26.89}{1.69} & \meanstd{32.89}{1.84}
            & \meanstd{38.89}{1.69} & \meanstd{11.36}{0.30}
            & \meanstd{0.415}{0.005} & \meanstd{27.00}{3.61} \\
        \midrule
        \multirow{4}{*}{\riboflowexp{}}
            & 50 & \meanstd{22.28}{0.63} & \meanstd{25.11}{0.42}
            & \meanstd{30.44}{0.54} & \meanstd{12.87}{0.24}
            & \meanstd{0.409}{0.004} & \meanstd{19.33}{4.04} \\
            & 100 & \meanstd{22.22}{0.54} & \meanstd{26.06}{0.59}
            & \meanstd{31.22}{0.98} & \meanstd{12.80}{0.41}
            & \meanstd{0.412}{0.003} & \meanstd{18.67}{1.53} \\
            & 200 & \meanstd{20.50}{0.73} & \meanstd{25.11}{0.42}
            & \meanstd{30.06}{0.69} & \meanstd{12.73}{0.22}
            & \meanstd{0.408}{0.003} & \meanstd{18.00}{2.65} \\
            & 400 & \meanstd{21.39}{0.54} & \meanstd{25.61}{0.92}
            & \meanstd{29.94}{0.77} & \meanstd{12.93}{0.42}
            & \meanstd{0.405}{0.004} & \meanstd{17.00}{2.00} \\
        \midrule
        \multirow{4}{*}{\riboflowmix{}}
            & 50 & \meanstd{29.00}{0.17} & \meanstd{29.67}{0.47}
            & \meanstd{37.28}{1.09} & \meanstd{11.26}{0.21}
            & \meanstd{0.428}{0.005} & \meanstd{23.00}{3.46} \\
            & 100 & \meanstd{29.06}{0.54} & \meanstd{33.67}{0.67}
            & \meanstd{38.44}{0.70} & \meanstd{11.48}{0.36}
            & \meanstd{0.452}{0.004} & \meanstd{22.33}{2.31} \\
            & 200 & \meanstd{27.61}{1.07} & \meanstd{30.89}{0.38}
            & \meanstd{36.89}{0.92} & \meanstd{11.52}{0.19}
            & \meanstd{0.443}{0.002} & \meanstd{21.67}{1.53} \\
            & 400 & \meanstd{28.39}{0.10} & \meanstd{30.78}{0.83}
            & \meanstd{37.39}{0.91} & \meanstd{10.99}{0.47}
            & \meanstd{0.437}{0.003} & \meanstd{20.67}{2.31} \\
        \midrule
        \multirow{4}{*}{\laribobase{}}
            & 50  & \meanstd{36.11}{2.07} & \meanstd{42.89}{1.73}
            & \meanstd{50.78}{2.00} & \meanstd{9.67}{0.60}
            & \meanstd{0.457}{0.012} & \meanstd{22.67}{1.53} \\
            & 100 & \meanstd{37.11}{0.98} & \meanstd{43.83}{0.50}
            & \meanstd{51.83}{1.20} & \meanstd{9.54}{0.43}
            & \meanstd{0.464}{0.007} & \meanstd{22.67}{1.15} \\
            & 200 & \meanstd{38.22}{1.27} & \meanstd{43.44}{0.48}
            & \meanstd{53.00}{0.67} & \meanstd{9.38}{0.29}
            & \meanstd{0.465}{0.006} & \meanstd{23.00}{1.73} \\
            & 400 & \meanstd{37.72}{1.13} & \meanstd{44.17}{0.60}
            & \meanstd{53.11}{1.55} & \meanstd{9.45}{0.36}
            & \meanstd{0.465}{0.007} & \meanstd{24.33}{1.15} \\
        \midrule
        \multirow{4}{*}{\laribotri{}}
            & 50  & \meanstd{37.11}{0.25} & \meanstd{45.22}{1.21}
            & \meanstd{54.06}{0.35} & \meanstd{8.80}{0.24}
            & \meanstd{0.474}{0.004} & \meanstd{29.00}{2.65} \\
            & 100 & \meanstd{37.78}{1.92} & \meanstd{46.89}{2.41}
            & \meanstd{56.89}{2.22} & \meanstd{8.56}{0.32}
            & \meanstd{0.487}{0.008} & \meanstd{28.33}{5.51} \\
            & 200 & \bestmeanstd{39.11}{2.12} & \bestmeanstd{48.39}{2.08}
            & \meanstd{58.83}{1.48} & \meanstd{8.59}{0.08}
            & \meanstd{0.491}{0.002} & \meanstd{27.67}{1.15} \\
            & 400 & \meanstd{38.61}{2.46} & \bestmeanstd{48.39}{2.75}
            & \bestmeanstd{59.33}{0.93} & \bestmeanstd{8.45}{0.02}
            & \bestmeanstd{0.494}{0.004} & \bestmeanstd{29.33}{3.79} \\
        \bottomrule
    \end{tabular}

    \vspace{1.0em}
    \textbf{(b) Codesignability}\\[0.45em]
    \begin{tabular}{lrrrrrrr}
        \toprule
        Method & Steps & \(\bothfour\,\uparrow\) & \(\bothfive\,\uparrow\)
        & \(\tmfortyfive\,\uparrow\) & scRMSD \(\downarrow\)
        & scTM \(\uparrow\) & \#Clust. \(\uparrow\) \\
        \midrule
        \multirow{4}{*}{\rnff{}}
            & 50 & -- & -- & -- & -- & -- & -- \\
            & 100 & -- & -- & -- & -- & -- & -- \\
            & 200 & -- & -- & -- & -- & -- & -- \\
            & 400 & -- & -- & -- & -- & -- & -- \\
        \midrule
        \multirow{4}{*}{\riboflowexp{}}
            & 50 & \meanstd{19.22}{0.35} & \meanstd{21.83}{0.76}
            & \meanstd{24.94}{0.10} & \meanstd{15.42}{0.21}
            & \meanstd{0.380}{0.003} & \meanstd{15.33}{2.52} \\
            & 100 & \meanstd{19.28}{0.42} & \meanstd{22.83}{1.09}
            & \meanstd{25.78}{1.25} & \meanstd{15.52}{0.11}
            & \meanstd{0.386}{0.001} & \meanstd{15.67}{2.08} \\
            & 200 & \meanstd{19.33}{1.61} & \meanstd{22.83}{1.76}
            & \meanstd{26.44}{2.55} & \meanstd{15.01}{0.14}
            & \meanstd{0.388}{0.006} & \meanstd{15.67}{1.15} \\
            & 400 & \meanstd{20.06}{0.19} & \meanstd{23.56}{0.48}
            & \meanstd{27.61}{0.69} & \meanstd{15.04}{0.42}
            & \meanstd{0.392}{0.003} & \meanstd{15.67}{1.53} \\
        \midrule
        \multirow{4}{*}{\riboflowmix{}}
            & 50 & \meanstd{26.61}{1.34} & \meanstd{27.89}{0.83}
            & \meanstd{31.83}{0.58} & \meanstd{14.01}{0.18}
            & \meanstd{0.419}{0.004} & \meanstd{19.00}{2.65} \\
            & 100 & \meanstd{26.67}{0.67} & \meanstd{30.00}{1.01}
            & \meanstd{32.50}{0.71} & \meanstd{13.61}{0.14}
            & \meanstd{0.437}{0.002} & \meanstd{19.33}{2.08} \\
            & 200 & \meanstd{26.00}{1.38} & \meanstd{29.06}{1.62}
            & \meanstd{33.06}{1.13} & \meanstd{14.15}{0.17}
            & \meanstd{0.399}{0.005} & \meanstd{19.00}{1.73} \\
            & 400 & \meanstd{26.61}{1.57} & \meanstd{27.83}{0.58}
            & \meanstd{34.72}{1.34} & \meanstd{13.01}{0.36}
            & \meanstd{0.403}{0.004} & \meanstd{19.33}{1.53} \\
        \midrule
        \multirow{4}{*}{\laribobase{}}
            & 50  & \meanstd{30.00}{2.77} & \meanstd{39.11}{2.38}
            & \meanstd{52.78}{2.18} & \meanstd{10.31}{0.29}
            & \meanstd{0.456}{0.010} & \meanstd{24.00}{1.00} \\
            & 100 & \meanstd{32.11}{2.14} & \meanstd{40.56}{1.87}
            & \meanstd{55.28}{1.55} & \meanstd{10.14}{0.13}
            & \meanstd{0.467}{0.007} & \meanstd{23.33}{3.06} \\
            & 200 & \bestmeanstd{32.56}{2.25} & \meanstd{41.28}{1.84}
            & \meanstd{56.72}{1.78} & \meanstd{9.85}{0.16}
            & \meanstd{0.473}{0.007} & \meanstd{25.33}{2.31} \\
            & 400 & \bestmeanstd{32.56}{1.58} & \bestmeanstd{41.72}{0.69}
            & \meanstd{57.17}{0.29} & \meanstd{9.93}{0.04}
            & \meanstd{0.475}{0.002} & \meanstd{25.67}{3.51} \\
        \midrule
        \multirow{4}{*}{\laribotri{}}
            & 50  & \meanstd{28.61}{1.13} & \meanstd{36.06}{1.93}
            & \meanstd{50.83}{2.09} & \meanstd{10.53}{0.28}
            & \meanstd{0.449}{0.009} & \bestmeanstd{28.00}{1.00} \\
            & 100 & \meanstd{30.78}{1.26} & \meanstd{39.00}{1.83}
            & \meanstd{56.11}{1.95} & \meanstd{9.86}{0.24}
            & \meanstd{0.471}{0.008} & \meanstd{25.33}{3.21} \\
            & 200 & \meanstd{32.06}{0.75} & \meanstd{39.94}{1.26}
            & \meanstd{57.89}{1.71} & \meanstd{9.67}{0.11}
            & \meanstd{0.477}{0.003} & \meanstd{27.33}{0.58} \\
            & 400 & \meanstd{32.33}{0.58} & \meanstd{40.17}{1.17}
            & \bestmeanstd{58.11}{0.84} & \bestmeanstd{9.55}{0.09}
            & \bestmeanstd{0.479}{0.003} & \bestmeanstd{28.00}{1.00} \\
        \bottomrule
    \end{tabular}
\end{table}
\FloatBarrier
\subsection{Triangle-update frequency across budgets}
\label{app:triangle_frequency}
We compare pair updates every two, three, and four blocks at the common sampling budgets. The two-block variant is \laribotri{}; \laribotrithree{} and \laribotrifour{} denote the three- and four-block variants. The representation, flow objective, and training corpus are shared; the number of refinement modules and parameter count vary with the update interval.

\begin{table}[!htbp]\centering
\caption{Designability by triangle-update interval. Mean $\pm$ standard deviation over three 600-RNA campaigns per setting; rates are percentages and scRMSD is in \AA{}.}
\vspace{0.4\baselineskip}
\label{tab:triangle_design_full}
\footnotesize\setlength{\tabcolsep}{2.5pt}\begin{tabular}{@{}lrrrrrrr@{}}\toprule
Model & Steps & $\bothfour\uparrow$ & $\bothfive\uparrow$ & $\tmfortyfive\uparrow$ & scRMSD $\downarrow$ & scTM $\uparrow$ & \#Clust.$\uparrow$\\\midrule
\multirow{4}{*}{\laribotri{}} & 50 & \meanstd{37.11}{0.25} & \meanstd{45.22}{1.21} & \meanstd{54.06}{0.35} & \meanstd{8.80}{0.24} & \meanstd{0.474}{0.004} & \meanstd{29.00}{2.65} \\
 & 100 & \meanstd{37.78}{1.92} & \meanstd{46.89}{2.41} & \meanstd{56.89}{2.22} & \meanstd{8.56}{0.32} & \meanstd{0.487}{0.008} & \meanstd{28.33}{5.51} \\
 & 200 & \meanstd{39.11}{2.12} & \meanstd{48.39}{2.08} & \meanstd{58.83}{1.48} & \meanstd{8.59}{0.08} & \meanstd{0.491}{0.002} & \meanstd{27.67}{1.15} \\
 & 400 & \meanstd{38.61}{2.46} & \meanstd{48.39}{2.75} & \meanstd{59.33}{0.93} & \meanstd{8.45}{0.02} & \meanstd{0.494}{0.004} & \meanstd{29.33}{3.79} \\
\midrule
\multirow{4}{*}{\laribotrithree{}} & 50 & \meanstd{31.17}{0.17} & \meanstd{37.67}{1.00} & \meanstd{49.00}{0.29} & \meanstd{9.54}{0.26} & \meanstd{0.459}{0.004} & \meanstd{26.33}{2.31} \\
 & 100 & \meanstd{36.50}{1.86} & \meanstd{42.00}{2.16} & \meanstd{52.83}{2.06} & \meanstd{9.06}{0.34} & \meanstd{0.479}{0.008} & \meanstd{26.33}{5.13} \\
 & 200 & \meanstd{38.50}{2.09} & \meanstd{44.00}{1.89} & \meanstd{56.00}{1.41} & \meanstd{8.64}{0.09} & \meanstd{0.487}{0.002} & \meanstd{26.33}{1.15} \\
 & 400 & \meanstd{38.00}{2.42} & \meanstd{45.50}{2.59} & \meanstd{55.33}{0.87} & \meanstd{8.40}{0.02} & \meanstd{0.489}{0.004} & \meanstd{27.33}{3.51} \\
\midrule
\multirow{4}{*}{\laribotrifour{}} & 50 & \meanstd{33.33}{0.29} & \meanstd{40.17}{1.09} & \meanstd{50.50}{0.33} & \meanstd{9.74}{0.26} & \meanstd{0.464}{0.004} & \meanstd{27.00}{2.65} \\
 & 100 & \meanstd{36.17}{1.84} & \meanstd{42.17}{2.17} & \meanstd{52.94}{1.94} & \meanstd{9.55}{0.35} & \meanstd{0.477}{0.008} & \meanstd{26.67}{5.03} \\
 & 200 & \meanstd{37.33}{2.03} & \meanstd{45.00}{1.92} & \meanstd{55.33}{1.36} & \meanstd{9.32}{0.09} & \meanstd{0.483}{0.002} & \meanstd{26.00}{1.00} \\
 & 400 & \meanstd{36.00}{2.29} & \meanstd{44.00}{2.50} & \meanstd{53.67}{0.83} & \meanstd{9.57}{0.03} & \meanstd{0.481}{0.004} & \meanstd{26.67}{3.51} \\
\bottomrule\end{tabular}\end{table}
\begin{table}[!htbp]\centering
\caption{Codesignability by triangle-update interval. Mean $\pm$ standard deviation over three 600-RNA campaigns per setting; rates are percentages and scRMSD is in \AA{}.}
\vspace{0.4\baselineskip}
\label{tab:triangle_codesign_full}
\footnotesize\setlength{\tabcolsep}{2.5pt}\begin{tabular}{@{}lrrrrrrr@{}}\toprule
Model & Steps & $\bothfour\uparrow$ & $\bothfive\uparrow$ & $\tmfortyfive\uparrow$ & scRMSD $\downarrow$ & scTM $\uparrow$ & \#Clust.$\uparrow$\\\midrule
\multirow{4}{*}{\laribotri{}} & 50 & \meanstd{28.61}{1.13} & \meanstd{36.06}{1.93} & \meanstd{50.83}{2.09} & \meanstd{10.53}{0.28} & \meanstd{0.449}{0.009} & \meanstd{28.00}{1.00} \\
 & 100 & \meanstd{30.78}{1.26} & \meanstd{39.00}{1.83} & \meanstd{56.11}{1.95} & \meanstd{9.86}{0.24} & \meanstd{0.471}{0.008} & \meanstd{25.33}{3.21} \\
 & 200 & \meanstd{32.06}{0.75} & \meanstd{39.94}{1.26} & \meanstd{57.89}{1.71} & \meanstd{9.67}{0.11} & \meanstd{0.477}{0.003} & \meanstd{27.33}{0.58} \\
 & 400 & \meanstd{32.33}{0.58} & \meanstd{40.17}{1.17} & \meanstd{58.11}{0.84} & \meanstd{9.55}{0.09} & \meanstd{0.479}{0.003} & \meanstd{28.00}{1.00} \\
\midrule
\multirow{4}{*}{\laribotrithree{}} & 50 & \meanstd{22.83}{0.93} & \meanstd{31.50}{1.69} & \meanstd{46.00}{1.89} & \meanstd{11.32}{0.30} & \meanstd{0.427}{0.009} & \meanstd{25.33}{1.15} \\
 & 100 & \meanstd{26.67}{1.09} & \meanstd{35.67}{1.69} & \meanstd{51.17}{1.77} & \meanstd{10.11}{0.25} & \meanstd{0.461}{0.008} & \meanstd{23.00}{3.00} \\
 & 200 & \meanstd{26.83}{0.60} & \meanstd{36.50}{1.15} & \meanstd{52.00}{1.53} & \meanstd{10.30}{0.12} & \meanstd{0.463}{0.003} & \meanstd{24.67}{0.58} \\
 & 400 & \meanstd{28.67}{0.50} & \meanstd{36.83}{1.09} & \meanstd{51.67}{0.76} & \meanstd{10.35}{0.09} & \meanstd{0.467}{0.003} & \meanstd{25.00}{1.00} \\
\midrule
\multirow{4}{*}{\laribotrifour{}} & 50 & \meanstd{22.00}{0.87} & \meanstd{32.83}{1.76} & \meanstd{48.83}{2.01} & \meanstd{11.84}{0.32} & \meanstd{0.436}{0.009} & \meanstd{27.00}{1.00} \\
 & 100 & \meanstd{23.83}{0.98} & \meanstd{34.50}{1.61} & \meanstd{50.33}{1.76} & \meanstd{11.45}{0.28} & \meanstd{0.449}{0.008} & \meanstd{22.67}{2.89} \\
 & 200 & \meanstd{25.17}{0.59} & \meanstd{36.67}{1.21} & \meanstd{52.67}{1.59} & \meanstd{11.01}{0.13} & \meanstd{0.458}{0.002} & \meanstd{25.00}{1.00} \\
 & 400 & \meanstd{25.50}{0.44} & \meanstd{37.00}{1.09} & \meanstd{53.50}{0.76} & \meanstd{10.89}{0.10} & \meanstd{0.463}{0.003} & \meanstd{25.67}{1.15} \\
\bottomrule\end{tabular}\end{table}
Across 50--400 steps, the two-block variant has higher strict codesignability than either sparser triangle-update variant. At 200 steps the gains are 5.23 and 6.89 percentage points. This pattern is stronger than the difference in strict designability. Base, reported in the main benchmark, remains competitive and shows that triangle refinement is not uniformly preferable for every endpoint.

\FloatBarrier
\section{Structural quality and novelty of generated structures}
\label{app:structural_quality}

\subsection{Secondary-structure agreement}
\label{app:secondary_agreement}

ViennaRNA RNAfold 2.4.14~\citep{lorenz2011viennarna} predicts a
minimum-free-energy base-pair set from each sequence. We compare it with two
annotations of the corresponding 3D structure: geometric contacts and DSSR
1.9.10~\citep{lu2015dssr}. Geometric AU, GC, and GU contacts require sequence
separation greater than three residues and both diagnostic atom distances to
be at most 4.0~\AA{}: A(N6)--U(O4) and A(N1)--U(N3); G(N1)--C(N3) and
G(O6)--C(N4); or G(N1)--U(O2) and G(O6)--U(N3), respectively. DSSR pairs are
restricted to canonical cis Watson--Crick/Watson--Crick AU/GC and GU wobble
interactions. This assay uses no learned refolding model.

\begin{table}[!htbp]
    \centering
    \caption{RNAfold agreement with geometric-contact and DSSR annotations on RNASolo dataset and 600 \laribotri{} structures. F1 and MCC are summarized per structure.}
    \vspace{0.4\baselineskip}
    \label{tab:rnafold_geometry_agreement}
    \small
    \setlength{\tabcolsep}{3.5pt}
    \begin{tabular}{llrrrr}
        \toprule
        3D annotation & Cohort & Mean F1 & Median F1 & Mean MCC & Median MCC \\
        \midrule
        Geometric contacts & Experimental & 0.5815 & 0.6341 & 0.5727 & 0.6320 \\
        & \laribotri{} & 0.5510 & 0.6023 & 0.5465 & 0.6051 \\
        \midrule
        DSSR & Experimental & 0.6451 & 0.7000 & 0.6367 & 0.7002 \\
        & \laribotri{} & 0.6056 & 0.6563 & 0.5919 & 0.6568 \\
        \bottomrule
    \end{tabular}
\end{table}

DSSR gives higher RNAfold agreement than geometric contacts in both
cohorts (\cref{tab:rnafold_geometry_agreement}). Under either annotation,
\laribotri{} closely approaches the experimental reference, with mean F1 and MCC
within 0.05 of the values for experimental structures.

\subsection{Atomistic geometry}
\label{app:atomistic}

We next evaluate local atomistic geometry directly from the generated
coordinates of 200-step 600-RNA campaigns of \laribobase{} and \laribotri{},
with \rnff{}, \riboflowmix{}, and the experimental and
synthetic subsets of the training corpus
as references. La-Ribo values are mean $\pm$ sample standard
deviation over three independently seeded campaigns; corpus
references are computed over all records of each subset.

\paragraph{Measured quantities.}
Clash counts steric violations per structure. Ribose pucker is assigned to
C3$'$-endo, C2$'$-endo, or a remaining category covering all other
conformations, and is reported as population fractions over valid residues. The
glycosidic torsion $\chi$ is classified as \emph{syn} or \emph{anti}, and
base-ring planarity is the RMSD of ring atoms from their least-squares plane.
Contact distances are averaged over the diagnostic atom pairs of canonical AU,
GC, and GU pairs, using the geometric criteria defined above. \rnff{} and
\riboflowmix{} generate backbone heavy atoms (frames plus
torsion-reconstructed phosphate and sugar atoms, including the glycosidic
nitrogen but no other base atoms), so $\chi$, planarity, and contact distances,
which require base atoms, are undefined for both; their clash counts cover these
backbone-level outputs only, whereas the \laribo{} variants are evaluated on all
heavy atoms.

\begin{table}[!htbp]
\centering
\caption{Atomistic geometry of generated structures at 200 steps
against the experimental (Exp) and synthetic (Syn) subsets of the
training corpus. La-Ribo values are mean $\pm$ sample standard deviation over
three 600-RNA campaigns; pucker and $\chi$ are population percentages over valid
residues; planarity and contact distances are in \AA{}.}
\vspace{0.4\baselineskip}
\label{tab:atomistic}
\scriptsize
\setlength{\tabcolsep}{2.5pt}
\begin{tabular}{lcccccc}
\toprule
Metric & Exp & Syn & \laribobase{} & \laribotri{} & \rnff{} & \riboflowmix{} \\
\midrule
Clash $\downarrow$
 & 0.089 & 0.646 & \meanstd{2.936}{0.200} & \meanstd{1.846}{0.097} & \meanstd{4.010}{0.261} & \meanstd{6.214}{0.283} \\
\midrule
C3$'$-endo (\%)
 & 83.482 & 86.167 & \meanstd{85.696}{0.061} & \meanstd{84.744}{0.067} & \meanstd{91.185}{0.094} & \meanstd{83.890}{0.168} \\
C2$'$-endo (\%)
 & 10.327 & 9.595 & \meanstd{8.667}{0.058} & \meanstd{9.648}{0.121} & \meanstd{6.466}{0.007} & \meanstd{2.636}{0.088} \\
Other pucker (\%)
 & 6.191 & 4.238 & \meanstd{5.636}{0.080} & \meanstd{5.608}{0.179} & \meanstd{2.349}{0.090} & \meanstd{13.474}{0.200} \\
\midrule
$\chi$ \emph{syn} (\%)
 & 4.358 & 3.136 & \meanstd{3.592}{0.031} & \meanstd{3.651}{0.026} & -- & -- \\
$\chi$ \emph{anti} (\%)
 & 95.642 & 96.864 & \meanstd{96.408}{0.031} & \meanstd{96.349}{0.026} & -- & -- \\
\midrule
Base-ring planarity RMSD
 & 0.0039 & 0.0022 & \meanstd{0.00113}{0.000} & \meanstd{0.00113}{0.000} & -- & -- \\
AU contact distance
 & 3.009 & 2.996 & \meanstd{3.100}{0.001} & \meanstd{3.089}{0.001} & -- & -- \\
GC contact distance
 & 3.060 & 3.074 & \meanstd{3.098}{0.003} & \meanstd{3.095}{0.001} & -- & -- \\
GU contact distance
 & 3.102 & 3.162 & \meanstd{3.151}{0.004} & \meanstd{3.147}{0.001} & -- & -- \\
\bottomrule
\end{tabular}
\end{table}

\paragraph{Steric quality and ribose pucker.}
Both \laribo{} variants produce fewer clashes than either baseline, and
\laribotri{} reduces the clash count by 37\% relative to \laribobase{}
(\cref{tab:atomistic}); this ordering differs from strict codesignability, where
Base is slightly ahead (\cref{tab:main}). The three-way pucker populations of
both \laribo{} variants track the experimental distribution in all categories.
\rnff{} concentrates 91.19\% of residues in C3$'$-endo against an experimental
83.48\%, consistent with a collapse toward A-form-like local geometry, while
\riboflowmix{} places almost no residues in C2$'$-endo (2.64\%
versus 10.33\%) and 13.47\% outside the two canonical pucker basins. Pucker is a
nucleotide-level property that the scaffold anchors do not determine, so this
agreement is attributable to the latent branch and the structured decoder
(\cref{app:vae_latent_controls}).

\paragraph{Glycosidic conformation and base-pair geometry.}
The \emph{syn} population is slightly below the experimental value
(3.59--3.65\% versus 4.36\%), indicating mild under-sampling of \emph{syn}
conformations. Base-ring planarity sits below the corpus references
(0.0011 versus 0.0022--0.0039), consistent with the idealized nucleotide templates of
the atom builder. Mean AU, GC,
and GU contact distances exceed the experimental values by 0.08--0.09,
0.03--0.04, and 0.05~\AA{}, respectively, across the two variants; the offsets
are one to two orders of magnitude larger than the across-campaign standard
deviations, indicating a systematic tendency toward slightly loose pairing. The synthetic and experimental subsets themselves differ measurably in these quantities (e.g., 0.646 versus 0.089 clashes per structure), and synthetic structures vastly outnumber experimental ones in the corpus (\cref{app:data_inventory}); the residual gap between generated and experimental geometry is therefore expected to partly reflect this distribution shift rather than model error alone.

\subsection{Novelty relative to the training corpus}
\label{app:novelty}

Our corpus assembles nearly all currently available high-quality RNA
structures, combining experimentally determined entries with synthetic ones (\cref{sec:data,app:data_inventory}). Under this coverage, the
usual novelty analysis, which asks whether a model generalizes beyond a standard
held-out split drawn from the same data sources, is no longer meaningful: such a
split would largely reproduce the training distribution rather than probe
extrapolation beyond it. We therefore measure directly how close generated
structures are to the training partition, and calibrate the measurement against
experimental structures that are known to be held out.

\paragraph{Protocol.}
TM-score is a length-normalized measure of global structural similarity taking
values in $[0,1]$, with higher values indicating closer folds; we compute it
with US-align~\citep{zhang2022us}. For each generated structure, the maximum
TM-score against the full training partition is denoted
$\mathrm{TM}_{\mathrm{train}}$, and $\mathrm{TM}_{\mathrm{exp}}$ is the
analogous maximum over all experimental structures in the corpus, excluding the
query structure itself. Sequence
identity and coverage are computed against the same nearest neighbour. A
structure is counted as a \emph{near-copy} when its nearest training neighbour
satisfies $\mathrm{TM} \geq 0.8$, sequence identity $\geq 0.8$, and coverage
$\geq 0.8$ in both directions. We evaluate 600-RNA campaigns from \laribobase{},
\laribotri{}, and \riboflowmix{} at 200 steps. Because
absolute TM values are difficult to interpret in isolation at this length range,
we apply the identical procedure to the 1,002 experimentally determined
structures in the cluster-held-out validation partition
(\cref{app:data_inventory}), whose MMseqs2 clusters are disjoint from all
training data; these structures bound what legitimate
generalization looks like under this measurement.

\begin{table}[!htbp]
\centering
\caption{Proximity to the training partition at 200 steps.
$\mathrm{TM}_{\mathrm{train}}$ and $\mathrm{TM}_{\mathrm{exp}}$ are maxima over
the training partition and over all experimental structures, respectively.
Near-copy requires $\mathrm{TM} \geq 0.8$, sequence identity $\geq 0.8$, and
coverage $\geq 0.8$ in both directions against the nearest training neighbour.
The validation cohort contains the 1,002 experimentally determined
structures held out from all training data and serves
as the reference cohort.}
\vspace{0.4\baselineskip}
\label{tab:novelty}
\small
\begin{tabular}{lcccc}
\toprule
Cohort & Mean $\mathrm{TM}_{\mathrm{train}}$ & $\mathrm{TM}_{\mathrm{train}} \geq 0.8$ (\%)
& Mean $\mathrm{TM}_{\mathrm{exp}}$ & Near-copy (\%) \\
\midrule
Validation partition (held-out) & 0.785 & 67.6 & 0.873 & 12.7 \\
\midrule
\riboflowmix{} & 0.541 & 20.6 & 0.573 & 14.7 \\
\laribobase{}  & 0.623 & 31.1 & 0.625 & 14.9 \\
\laribotri{}   & 0.644 & 37.8 & 0.659 & 16.1 \\
\bottomrule
\end{tabular}
\end{table}

\paragraph{Proximity to the training partition.}
Held-out experimental structures have mean $\mathrm{TM}_{\mathrm{train}} =
0.785$, and 67.6\% of them have a training neighbour at $\mathrm{TM} \geq 0.8$
(\cref{tab:novelty}): sequence-level clustering does not produce structural
separation at this length range. Both \laribo{} variants are substantially less
similar to the training partition than real held-out structures are (31.1\% and
37.8\%), so the generated distributions are not concentrated on reproductions of
training examples. The generated cohorts also score similarly against
experimental and training references, with gaps of $+0.002$, $+0.015$, and
$+0.032$ for Base, Tri, and \riboflowmix{}, and \laribotri{} reaches the highest
$\mathrm{TM}_{\mathrm{exp}}$ among them (\cref{tab:novelty}). The near-copy
rate additionally requires sequence identity of at least 0.8 to a training
entry; the cluster-based split excludes cross-split matches above 90\% sequence identity at 95\% coverage, but not matches in the 80--90\% range, so held-out structures can still satisfy this condition. Their
near-copy rate (12.7\%) is close to those of the generated cohorts
(14.7--16.1\%), so the near-copy assessment agrees with the structural
columns above.

\section{Sampling sensitivity and computational cost}
\label{app:generation_landscape}
\subsection{Modality-schedule sensitivity}
\label{app:schedule_sensitivity}
We vary the scaffold and latent schedule powers independently to characterize sensitivity, separately from the fixed main-benchmark configuration. The balanced sensitivity collection covers four architectures, three step budgets (100, 200, 400), $\rho_B\in\{0.50,0.70,0.85,1.00,1.25\}$, and $\rho_Z\in\{1.00,1.20,1.40,1.80,2.00\}$. Across the complete balanced collection, every operating point contains 600 generated RNAs. The powers control progress along the two probability paths; they are schedule parameters, not multiplicative scales on the Gaussian noise.

The heatmaps visualize the observed response envelope across the sensitivity collection, while the marginal summaries average over the remaining factors. Neither display defines the main-benchmark protocol.
\begin{figure}[!htbp]
    \centering
    \includegraphics[page=1,width=0.95\textwidth]{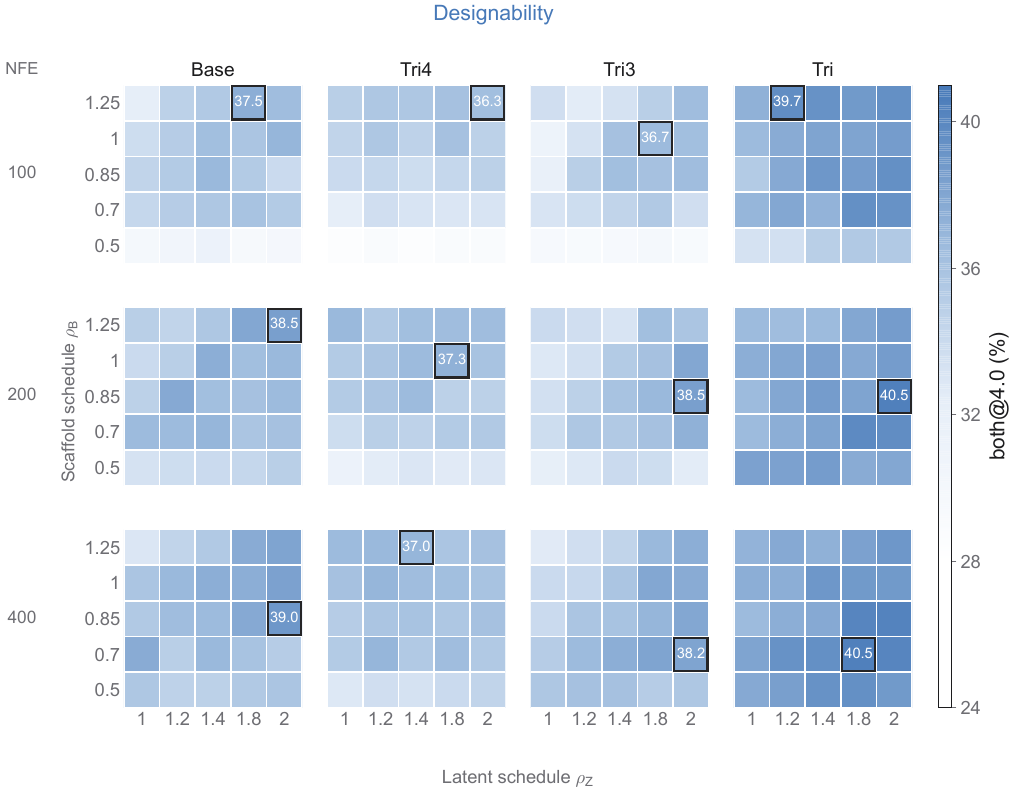}
    \caption{Strict designability across modality schedules and four architectures. Rows show 100, 200, and 400 steps; each schedule-pair cell contains 600 RNAs. Panels share a color scale, and black outlines mark maxima.}
    \label{fig:generation_landscape_design}
\end{figure}
\begin{figure}[!htbp]
    \centering
    \includegraphics[page=2,width=0.95\textwidth]{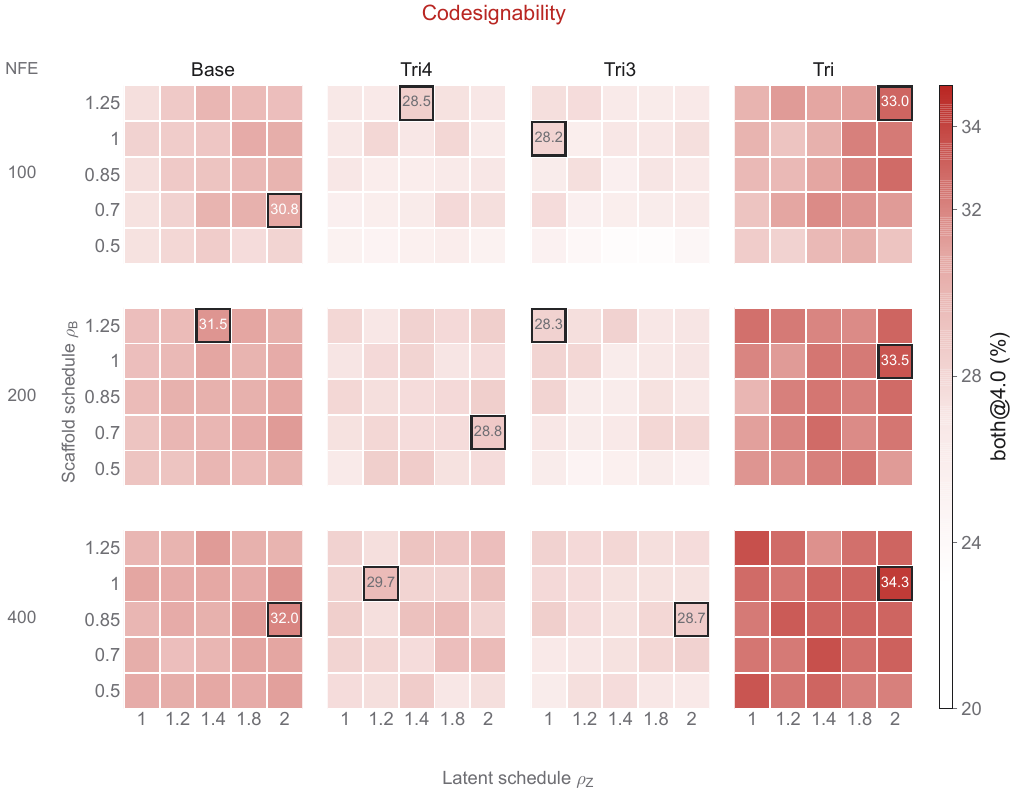}
    \caption{Strict codesignability on the schedule grid of \cref{fig:generation_landscape_design}. Panels share a color scale, and black outlines mark maxima.}
    \label{fig:generation_landscape_codesign}
\end{figure}
\subsection{Sampling-factor response and runtime}
\label{app:sampling_factors}

We average each factor over the full balanced sensitivity collection and the remaining factor levels.
\Cref{tab:sampling_factors} reports the architecture-averaged marginal responses. The architecture-specific curves are shown in main-text \cref{fig:sampling_factor_response_main}, while the heatmaps above retain interactions averaged out in these summaries.

\begin{table}[!htbp]
    \centering
    \caption{Marginal strict $\bothfour$ success by sampling factor, averaged over the remaining grid settings. Each setting has 600 RNAs; rates are percentages.}
    \vspace{0.4\baselineskip}
    \label{tab:sampling_factors}
    \footnotesize
    \begin{tabular}{llrr}
        \toprule
        Factor & Level & Designability \(\uparrow\)
        & Codesignability \(\uparrow\) \\
        \midrule
        Solver steps \(K\) & 100 & 31.90 & 25.01 \\
        & 200 & 33.18 & 26.07 \\
        & 400 & 33.84 & 26.67 \\
        \midrule
        Scaffold power \(\rho_B\) & 0.50 & 31.03 & 25.08 \\
        & 0.70 & 33.23 & 25.82 \\
        & 0.85 & 33.48 & 26.04 \\
        & 1.00 & 33.59 & 26.24 \\
        & 1.25 & 33.53 & 26.40 \\
        \midrule
        Latent power \(\rho_Z\) & 1.00 & 32.14 & 25.72 \\
        & 1.20 & 32.69 & 25.77 \\
        & 1.40 & 32.99 & 25.94 \\
        & 1.80 & 33.47 & 26.06 \\
        & 2.00 & 33.57 & 26.08 \\
        \bottomrule
    \end{tabular}
\end{table}

Increasing the budget from 100 to 400 steps changes either marginal strict endpoint by less than two percentage points. Sensitivity flattens at scaffold powers 0.70--1.25 and latent powers 1.80--2.00, although some architecture-specific maxima remain at the upper latent-power boundary. The sampled range thus contains a broad high-performing region without establishing an interior optimum.

\begin{figure}[!htbp]
    \centering
    \includegraphics[width=\linewidth]{%
        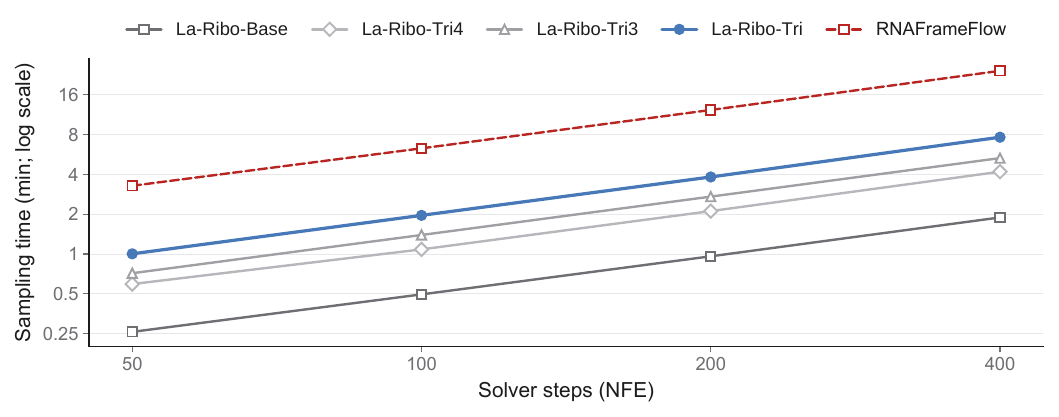}
    \caption{Generation time for 600 RNAs by solver steps, excluding inverse folding, refolding, and scoring. La-Ribo uses eight A800 GPUs; \rnff{} time is throughput-normalized. Timing scopes differ; steps are categorical and time is logarithmic.}
    \label{fig:sampling_runtime}
\end{figure}

Base has the lowest measured sampling cost; Tri costs approximately four times as much, with \laribotrithree{} and \laribotrifour{} between them. The measured \rnff{} pipeline costs about three times as much as Tri. Different timing scopes make these implementation-level comparisons rather than intrinsic algorithmic speedups.

\FloatBarrier
\section{Evaluation protocols, robustness, and inverse folding}
\label{app:evaluation_robustness}

\subsection{Primary evaluation settings and baseline implementation}
\label{app:evaluation_protocols}

\paragraph{Unconditional generation.}
Each campaign contains 600 structures, with 50 samples at each length from
40 to 150 nt in increments of ten. Three independently seeded campaigns are
evaluated for every main-benchmark model and sampling budget, and reported
standard deviations summarize sampling variation of the fixed trained models.
Designability uses eight inverse-folded candidate sequences per generated
structure, whereas codesignability evaluates the emitted sequence.
All reported generated structures are direct sampler outputs.

\paragraph{Inverse folding and refolding.}
We use the released \grnade{} ARv1 single-state all-data weights at
temperature 0.1. \rhofold{} receives each candidate sequence as its sole
alignment input, without external MSA search, and its output coordinates are
unrelaxed. These settings are shared across methods. Strict designability
requires at least one of the eight candidates to satisfy both thresholds;
continuous best-candidate metrics and cluster counts follow the same
definitions throughout the expanded result tables.

\paragraph{Structural cluster count.}
For each 600-RNA campaign, the eligible set consists of generated
backbones with scTM$\geq0.45$ under the relevant evaluation route: the maximum
across eight \grnade{} sequences for designability, or the emitted sequence for
codesignability. We cluster the original generated backbone PDBs, not the
\rhofold{} refolds, with qTMclust~\citep{zhang2022us} (\texttt{-TMcut 0.45})
and count the output clusters as \#Clust. We cluster the three campaigns
independently and report the mean $\pm$ sample standard deviation of their
counts. Failed or missing samples do not enter clustering but remain in the
600-sample success-rate denominator.

\Needspace{7\baselineskip}

\subsection{Unconditional generation with an alternative refolding model}
\label{app:opendde_robustness}
OpenDDE~\citep{aureka2026opendde} replaces RhoFold while retaining each generated structure, candidate-sequence route, and score definition. \Cref{tab:main_opendde} reports this evaluation. Both La-Ribo variants retain higher strict designability and codesignability than the corresponding baselines.
\begin{table}[!t]
    \centering
    \caption{Unconditional generation with OpenDDE instead of RhoFold under the settings of \cref{tab:main}. Rates are percentages; scRMSD is in \AA{}, and bold marks the best mean per panel.}
    \vspace{0.4\baselineskip}
    \label{tab:main_opendde}
    \footnotesize
    \setlength{\tabcolsep}{2.2pt}
    \textbf{(a) Designability}\\[0.45em]
    \begin{tabular}{lrrrrrrr}
        \toprule
        Method & Steps & \(\bothfour\,\uparrow\) & \(\bothfive\,\uparrow\)
        & \(\tmfortyfive\,\uparrow\) & scRMSD \(\downarrow\)
        & scTM \(\uparrow\) & \#Clust. \(\uparrow\) \\
        \midrule
        \multirow{4}{*}{\rnff{}}
            & 50 & \meanstd{25.39}{1.64} & \meanstd{30.22}{1.36}
            & \meanstd{39.39}{0.35} & \meanstd{12.33}{0.49}
            & \meanstd{0.426}{0.006} & \meanstd{31.67}{4.04} \\
            & 100 & \meanstd{27.06}{0.51} & \meanstd{33.44}{0.92}
            & \meanstd{42.83}{1.17} & \meanstd{12.10}{0.21}
            & \meanstd{0.434}{0.003} & \meanstd{33.33}{4.04} \\
            & 200 & \meanstd{26.61}{1.58} & \meanstd{33.56}{1.95}
            & \meanstd{42.11}{3.06} & \meanstd{12.28}{0.40}
            & \meanstd{0.429}{0.009} & \meanstd{35.00}{2.65} \\
            & 400 & \meanstd{27.50}{0.29} & \meanstd{33.33}{0.87}
            & \meanstd{41.56}{0.59} & \meanstd{12.08}{0.31}
            & \meanstd{0.429}{0.005} & \meanstd{32.67}{2.08} \\
        \midrule
        \multirow{4}{*}{\riboflowexp{}}
            & 50 & \meanstd{22.67}{0.67} & \meanstd{27.67}{0.50}
            & \meanstd{34.00}{1.20} & \meanstd{12.29}{0.30}
            & \meanstd{0.450}{0.002} & \meanstd{26.33}{2.52} \\
            & 100 & \meanstd{22.94}{1.92} & \meanstd{28.89}{2.20}
            & \meanstd{36.11}{0.86} & \meanstd{12.04}{0.35}
            & \meanstd{0.456}{0.005} & \meanstd{28.33}{2.89} \\
            & 200 & \meanstd{22.78}{0.38} & \meanstd{28.56}{0.38}
            & \meanstd{36.00}{0.33} & \meanstd{12.29}{0.21}
            & \meanstd{0.452}{0.003} & \meanstd{26.33}{4.04} \\
            & 400 & \meanstd{23.33}{0.17} & \meanstd{28.44}{0.42}
            & \meanstd{34.83}{0.87} & \meanstd{12.41}{0.31}
            & \meanstd{0.447}{0.002} & \meanstd{23.67}{0.58} \\
        \midrule
        \multirow{4}{*}{\riboflowmix{}}
            & 50 & \meanstd{29.22}{0.61} & \meanstd{32.28}{0.57}
            & \meanstd{41.06}{0.93} & \meanstd{11.11}{0.26}
            & \meanstd{0.489}{0.003} & \meanstd{30.00}{3.00} \\
            & 100 & \meanstd{29.89}{1.46} & \meanstd{34.78}{1.94}
            & \meanstd{42.56}{1.21} & \meanstd{10.85}{0.31}
            & \meanstd{0.478}{0.004} & \meanstd{32.00}{2.65} \\
            & 200 & \meanstd{30.00}{1.15} & \meanstd{34.17}{0.44}
            & \meanstd{43.89}{0.82} & \meanstd{10.92}{0.25}
            & \meanstd{0.499}{0.004} & \meanstd{30.33}{3.21} \\
            & 400 & \meanstd{30.50}{0.58} & \meanstd{33.56}{0.35}
            & \meanstd{41.67}{1.18} & \meanstd{10.84}{0.34}
            & \meanstd{0.483}{0.003} & \meanstd{27.33}{1.15} \\
        \midrule
        \multirow{4}{*}{\laribobase{}}
            & 50 & \meanstd{34.33}{1.64} & \meanstd{45.78}{1.06}
            & \meanstd{58.22}{0.42} & \meanstd{9.61}{0.33}
            & \meanstd{0.492}{0.007} & \meanstd{32.00}{3.61} \\
            & 100 & \meanstd{33.11}{2.20} & \meanstd{45.11}{0.35}
            & \meanstd{56.72}{0.54} & \meanstd{9.70}{0.34}
            & \meanstd{0.490}{0.006} & \meanstd{29.67}{2.08} \\
            & 200 & \meanstd{34.11}{0.79} & \meanstd{46.89}{1.34}
            & \meanstd{59.44}{1.84} & \meanstd{9.56}{0.31}
            & \meanstd{0.495}{0.009} & \meanstd{31.33}{1.53} \\
            & 400 & \meanstd{33.78}{1.84} & \meanstd{46.44}{1.78}
            & \meanstd{60.06}{2.11} & \meanstd{9.45}{0.42}
            & \meanstd{0.495}{0.008} & \meanstd{33.00}{2.00} \\
        \midrule
        \multirow{4}{*}{\laribotri{}}
            & 50 & \meanstd{37.50}{1.74} & \meanstd{46.72}{0.67}
            & \meanstd{59.67}{0.44} & \meanstd{8.93}{0.37}
            & \meanstd{0.517}{0.001} & \meanstd{41.33}{3.06} \\
            & 100 & \meanstd{39.72}{2.56} & \meanstd{49.50}{2.20}
            & \meanstd{63.50}{1.01} & \bestmeanstd{8.56}{0.50}
            & \meanstd{0.531}{0.008} & \meanstd{41.00}{2.65} \\
            & 200 & \meanstd{38.33}{0.76} & \meanstd{49.72}{0.42}
            & \meanstd{63.94}{0.54} & \meanstd{8.82}{0.10}
            & \meanstd{0.530}{0.004} & \meanstd{39.67}{1.53} \\
            & 400 & \bestmeanstd{39.83}{1.00} & \bestmeanstd{50.28}{0.67}
            & \bestmeanstd{65.22}{0.67} & \bestmeanstd{8.56}{0.05}
            & \bestmeanstd{0.537}{0.001} & \bestmeanstd{43.33}{8.02} \\
        \bottomrule
    \end{tabular}

    \vspace{0.7em}
    \textbf{(b) Codesignability}\\[0.45em]
    \begin{tabular}{lrrrrrrr}
        \toprule
        Method & Steps & \(\bothfour\,\uparrow\) & \(\bothfive\,\uparrow\)
        & \(\tmfortyfive\,\uparrow\) & scRMSD \(\downarrow\)
        & scTM \(\uparrow\) & \#Clust. \(\uparrow\) \\
        \midrule
                \multirow{4}{*}{\riboflowexp{}}
            & 50 & \meanstd{20.89}{0.35} & \meanstd{25.72}{1.64}
            & \meanstd{30.50}{1.59} & \meanstd{15.44}{0.09}
            & \meanstd{0.431}{0.005} & \meanstd{18.00}{2.00} \\
            & 100 & \meanstd{20.61}{0.86} & \meanstd{26.11}{0.84}
            & \meanstd{32.00}{0.76} & \meanstd{15.12}{0.38}
            & \meanstd{0.437}{0.004} & \meanstd{20.33}{1.15} \\
            & 200 & \meanstd{20.72}{1.84} & \meanstd{25.89}{1.08}
            & \meanstd{32.17}{0.29} & \meanstd{15.25}{0.09}
            & \meanstd{0.433}{0.001} & \meanstd{19.33}{0.58} \\
            & 400 & \meanstd{21.11}{1.93} & \meanstd{26.89}{0.42}
            & \meanstd{32.39}{0.67} & \meanstd{15.07}{0.39}
            & \meanstd{0.436}{0.001} & \meanstd{18.33}{3.79} \\
        \midrule
        \multirow{4}{*}{\riboflowmix{}}
            & 50 & \meanstd{27.67}{1.66} & \meanstd{31.83}{1.52}
            & \meanstd{38.11}{0.86} & \meanstd{14.19}{0.12}
            & \meanstd{0.479}{0.004} & \meanstd{21.67}{2.31} \\
            & 100 & \meanstd{27.50}{1.36} & \meanstd{30.56}{0.91}
            & \meanstd{38.72}{0.19} & \meanstd{14.11}{0.33}
            & \meanstd{0.458}{0.005} & \meanstd{24.00}{1.73} \\
            & 200 & \meanstd{27.67}{1.34} & \meanstd{29.06}{1.16}
            & \meanstd{39.22}{1.35} & \meanstd{13.79}{0.11}
            & \meanstd{0.435}{0.002} & \meanstd{23.00}{1.00} \\
            & 400 & \meanstd{27.78}{0.79} & \meanstd{32.00}{0.47}
            & \meanstd{38.89}{0.93} & \meanstd{13.14}{0.42}
            & \meanstd{0.491}{0.002} & \meanstd{22.00}{3.00} \\
        \midrule
        \multirow{4}{*}{\laribobase{}}
            & 50 & \meanstd{31.17}{2.89} & \meanstd{40.72}{2.38}
            & \meanstd{56.28}{1.84} & \meanstd{10.76}{0.31}
            & \meanstd{0.500}{0.011} & \meanstd{30.33}{0.58} \\
            & 100 & \meanstd{32.00}{1.42} & \meanstd{42.78}{2.20}
            & \meanstd{58.22}{0.75} & \meanstd{10.58}{0.51}
            & \meanstd{0.508}{0.010} & \meanstd{32.00}{3.00} \\
            & 200 & \meanstd{32.50}{1.26} & \meanstd{44.44}{2.01}
            & \meanstd{59.78}{1.99} & \meanstd{10.45}{0.52}
            & \meanstd{0.514}{0.012} & \meanstd{32.33}{1.53} \\
            & 400 & \bestmeanstd{33.72}{2.26} & \bestmeanstd{44.78}{1.70}
            & \meanstd{58.83}{1.73} & \bestmeanstd{10.44}{0.25}
            & \meanstd{0.512}{0.007} & \meanstd{29.00}{1.00} \\
        \midrule
        \multirow{4}{*}{\laribotri{}}
            & 50 & \meanstd{28.33}{1.50} & \meanstd{36.94}{1.29}
            & \meanstd{56.11}{2.12} & \meanstd{11.73}{0.25}
            & \meanstd{0.501}{0.007} & \meanstd{40.00}{2.00} \\
            & 100 & \meanstd{29.39}{0.95} & \meanstd{38.39}{0.86}
            & \meanstd{58.33}{1.04} & \meanstd{11.15}{0.30}
            & \meanstd{0.518}{0.006} & \meanstd{36.67}{3.79} \\
            & 200 & \meanstd{33.33}{2.05} & \meanstd{41.00}{2.20}
            & \meanstd{60.22}{1.78} & \meanstd{11.17}{0.14}
            & \meanstd{0.522}{0.007} & \meanstd{40.00}{2.65} \\
            & 400 & \meanstd{32.06}{1.23} & \meanstd{41.56}{1.42}
            & \bestmeanstd{61.00}{2.42} & \meanstd{11.27}{0.05}
            & \bestmeanstd{0.526}{0.004} & \bestmeanstd{41.00}{4.36} \\
        \bottomrule
    \end{tabular}
\end{table}
\subsection{Scaffold-conditioned inverse-folding protocol}
\label{app:conditional_sequence_protocol}

\paragraph{Targets.}
The validation partition is held out from all training stages by sequence cluster (\cref{app:data_inventory}); its complete experimental subset comprises 1,002 records (RNASolo and RhoFold-PDB chains) spanning 107 clusters, and every record serves as a scaffold target.

\paragraph{Inputs and scoring.}
La-Ribo fixes the centered P/C4$'$/N scaffold and its clean time endpoint, then integrates the latent branch for 200 steps. \riboflowmix{} fixes native residue frames while evolving sequence states. Neither joint model receives the native sequence. Dedicated inverse folders, gRNAde and NA-MPNN, receive canonicalized native structures. Each method supplies eight candidates per target, for 8,016 candidate sequences. RhoFold refolding and C4$'$-based scoring use the same thresholds as the unconditional benchmark. Mean-of-eight metrics average candidate scores per target. Best-of-eight strict success requires a single candidate satisfying both thresholds. Best scRMSD and best scTM are optimized separately within a target and then averaged. All 1,002 targets and 8,016 candidate slots remain in success-rate denominators; nonfinite predictions count as failures. Continuous means use the finite predictions. Each method is evaluated with three independently seeded runs, and reported values are the mean $\pm$ sample standard deviation across runs. The complete inverse-folding comparison is reported in \cref{tab:conditional_sequence_main}.

\FloatBarrier
\section{Limitations and broader impact}
\label{app:broader_impacts}

\paragraph{Limitations.}
Experimentally determined RNA structures are scarce, so most of our corpus
consists of predicted or synthetic structures, whose atom-level geometry does
not fully match that of experimentally resolved structures
(\cref{app:atomistic}). This distribution gap limits the atomistic accuracy that
models trained on this corpus can attain. We expect it to gradually narrow as
structure prediction models continue to improve. Besides, long RNAs remain difficult because high-quality training structures at
longer lengths are scarce in both experimental and synthetic sources. Only
0.31\% of the corpus exceeds 160 nt: the experimental RNASolo and RhoFold
subsets allocate just 1.18\% and 1.90\% of their entries to this range, and
OpenFold3, which dominates the corpus, contributes only 0.06\%
(\cref{fig:training_length_distributions}). Consistent with this thinning
coverage, strict success declines toward the upper end of the 40--150-nt
benchmark range (\cref{fig:length_resolved}). Because the missing supervision
is a property of the available data rather than of the model, we expect this
gap to persist until larger collections of high-quality long-RNA structures,
experimental or predicted, become available.

\paragraph{Broad impact.}
La-Ribo offers a framework for joint RNA sequence and structure generation. We hope it will support future RNA research by enabling studies of sequence--structure relationships and informing new RNA design methods. Functional applications should undergo experimental validation and biosafety review because generated sequences may have unintended biological effects.

\end{document}